\documentclass{JHEP3} 

\usepackage{epsfig,multicol,bbm}

\newcommand\fverb{\setbox\fverbbox=\hbox\bgroup\verb}
\newcommand\fverbdo{\egroup\medskip\noindent%
			\fbox{\unhbox\fverbbox}\ }
\newcommand\fverbit{\egroup\item[\fbox{\unhbox\fverbbox}]}
\newbox\fverbbox

\title{Likelihood analysis of the next-to-minimal supergravity motivated model}

\author{Csaba Bal\'azs \\ 
	School of Physics, Monash University, Melbourne Victoria 3800, Australia \\
	E-mail: \email{csaba.balazs@sci.monash.edu.au}}

\author{Daniel Carter \\
	School of Physics, Monash University, Melbourne Victoria 3800, Australia \\
	E-mail: \email{daniel.carter@sci.monash.edu.au}}

\received{\today} 		
\accepted{\today}		

\abstract{
In anticipation of data from the Large Hadron Collider (LHC) and the potential discovery of supersymmetry, in this work we seek an answer to the following: What are the chances that supersymmetry will be found at the LHC?  Will the LHC data be enough to discover a given supersymmetric model?  And what other measurements can assist the LHC establish the presence of supersymmetry?  
As a step toward answering these general questions, we calculate the odds of the next-to-minimal version of the popular supergravity motivated model (NmSuGra) being discovered at the LHC to be 4:3 (57 \%).
We also demonstrate that viable regions of the NmSuGra parameter space outside the LHC reach can be covered by upgraded versions of dark matter direct detection experiments, such as super-CDMS, at 99 \% confidence level.  
Due to the similarities of the models, we expect very similar results for the constrained minimal supersymmetric standard model (CMSSM).
}


\begin{document} 


\section{Introduction}


Supersymmetry is one of the most robust theories that can solve outstanding problems of the standard model (SM) of elementary particles.  The theory naturally explains the dynamics of electroweak symmetry breaking while preserving the hierarchy of fundamental energy scales. It also readily accommodates dark matter, the asymmetry between baryons and anti-baryons, the unification of gauge forces, gravity, and more.
%
%
But if supersymmetry is the solution to the problems of the standard model, then its natural scale is the electroweak scale, and it is expected to be observed in upcoming experiments, most notably the CERN Large Hadron Collider (LHC).  In this work, we will attempt to determine, quantitatively, what the chances are that this may occur for the simplified case of a constrained supersymmetric model.


One of the main motivations for supersymmetry is that it can naturally bridge 
the hierarchy between the weak and Planck scales.  Unfortunately, the presence 
of the superpotential $\mu$ term in the minimal supersymmetric extension of the 
standard model (MSSM) undermines this very aim \cite{Kim:1983dt}.  Experimental 
data have also squeezed the MSSM into fine-tuned regions, creating the 
supersymmetric little hierarchy problem\cite{Giudice:2008bi}.
%
%
Extensions of the MSSM by gauge singlet superfields not only resolve the $\mu$ 
problem, but can also ameliorate the little hierarchy problem \cite{Dermisek:2005ar, 
BasteroGil:2000bw, Gunion:2008kp}.  In the next-to-minimal MSSM (NMSSM), the 
$\mu$ term is dynamically generated and no dimensionful parameters are 
introduced in the superpotential (other than the vacuum expectation values that 
are all naturally weak scale), making the NMSSM a truly natural model 
\cite{Fayet:1974pd, Nilles:1982dy, Frere:1983ag, Derendinger:1983bz, 
Greene:1986th, Ellis:1986mq, Durand:1988rg, Drees:1988fc, Ellis:1988er, 
Pandita:1993tg, Pandita:1993hx, Ellwanger:1993xa, Ananthanarayan:1995zr,
Ananthanarayan:1995xq, Ananthanarayan:1996zv, Ellwanger:1996gw, Elliott:1994ht, 
King:1995vk}.  


Over the last two decades, due to its simplicity and elegance, the constrained 
MSSM (CMSSM) and the minimal supergravity-motivated (mSuGra) model became a 
standard in supersymmetry phenomenology.  Guided by this, within the NMSSM, we 
impose the universality of sparticle masses, gaugino masses, and tri-linear 
couplings at the grand unification theory (GUT) scale, thereby defining the
next-to-minimal supergravity-motivated (NmSuGra) model.
%
%
This approach ensures that all dimensionful parameters of the NMSSM scalar 
potential also naturally arise from supersymmetry breaking in a minimal fashion.  
NmSuGra also reduces the electroweak and dark matter fine-tunings of mSuGra.


Using a Bayesian likelihood analysis, we identify the regions in the parameter space of the NmSuGra model that are preferred by the present experimental limits from various collider, astrophysical, and low-energy measurements.  We combine theoretical exclusions with experimental limits from the CERN Large Electron-Positron (LEP) collider, the Fermilab Tevatron, NASA's Wilkinson Microwave Anisotropy Probe (WMAP) satellite (and other related astrophysical measurements), the Soudan Cryogenic Dark Matter Search (CDMS), the Brookhaven Muon g$-$2 Experiment, and various b-physics measurements including the rare decay branching fractions $b \to s \gamma$ and $B_s \to l^+ l^-$. 
%
%
Thus we show that, given current experimental constraints, the favored parameter space can be detected by a combination of the LHC and an upgraded CDMS at the 95 \% confidence level.

In the next section we define the next-to-minimal version of the supergravity motivated model (NmSuGra).  Then, in Section \ref{sec:Bayes}, we summarize the main concepts of Bayesian inference that we use in this work.  Section \ref{sec:likelihood} contains the numerical results of our likelihood analysis, and Section \ref{sec:detection} gives the outlook for the experimental detection of NmSuGra.

\section{The next-to-minimal supergravity motivated model}
\label{sec:NmSuGra}


The next-to-minimal supersymmetric model (NMSSM) is defined by the superpotential 
\begin{eqnarray}
 W_{NMSSM} = W_{MSSM}|_{\mu = 0} + \lambda \hat{S} \hat{H}_u \cdot \hat{H}_d + \frac{\kappa}{3} \hat{S}^3,
\label{eq:W_NMSSM}
\end{eqnarray}
where 
\begin{eqnarray}
 W_{MSSM}|_{\mu = 0} = 
  y_t    \hat{Q}_L \cdot \hat{H}_u \hat{T}_R^c -
  y_b    \hat{Q}_L \cdot \hat{H}_d \hat{B}_R^c -
  y_\tau \hat{L}_L \cdot \hat{H}_d \hat{L}_R^c
\label{eq:W_MSSM}
\end{eqnarray}
is the MSSM superpotential containing only Yukawa terms and having $\mu$ set to zero \cite{Ellwanger:2005dv}.  The left (right) handed matter superfields $\hat{Q}_L$, $\hat{L}_L$ ($\hat{T}_R$, $\hat{B}_R$, $\hat{L}_R$), and Higgs $\hat{H}_{u,d}$ superfields are $SU(2)_L$ doublets (singlets), while $\hat{S}$ is a standard gauge singlet.  The couplings $\lambda$, $\kappa$, and $y_i$ are dimensionless, and $\hat{X} \cdot \hat{Y} = \epsilon_{\alpha\beta} \hat{X}^\alpha \hat{Y}^\beta$ with the fully antisymmetric tensor normalized as $\epsilon_{11} = 1$.  The corresponding soft supersymmetry breaking terms are
\begin{eqnarray}
 \mathcal{L}^{soft}_{NMSSM} = && \mathcal{L}^{soft}_{MSSM}|_{B = 0} - M_S^2 |\tilde{S}|^2 - (\lambda A_\lambda \tilde{S} H_u \cdot H_d + \frac{\kappa A_\kappa}{3} \tilde{S}^3 + h.c.) .
\end{eqnarray}
Here 
\begin{eqnarray}
 \mathcal{L}^{soft}_{MSSM}|_{B = 0} &=& \mathcal{L}^{soft}_{gaugino} + \mathcal{L}^{soft}_{scalar} + \mathcal{L}^{soft}_{tri-linear} ,
\end{eqnarray}
contains the mass terms for the twelve gauginos ($i=1...3$, $a=1...8$)
\begin{eqnarray}
 \mathcal{L}^{soft}_{gaugino} &=& -\frac{1}{2} (M_1 \bar{\tilde{B}} \tilde{B} + M_2 \bar{\tilde{W}}_i \tilde{W}_i + M_3 \bar{\tilde{G}}_a \tilde{G}_a + h.c) ,
\end{eqnarray}
the sfermions and Higgses  
\begin{eqnarray}
 \mathcal{L}^{soft}_{scalar} = - &(&M_{Q}^2 |\tilde{Q}|^2 + M_{T_R}^2 |\tilde{T}_R|^2 + M_{B_R}^2 |\tilde{B}_R|^2 + M_{L}^2 |\tilde{L}|^2 + M_{L_R}^2 |\tilde{L}_R|^2 + \nonumber \\ &&M_{H_u}^2 |H_u|^2 + M_{H_d}^2 |H_d|^2 ) ,
\end{eqnarray}
and the soft tri-linear terms
\begin{eqnarray}
 \mathcal{L}^{soft}_{tri-linear} &=& - (y_t A_t \tilde{Q} \cdot H_u \tilde{T}_R^c - y_b A_b \tilde{Q} \cdot H_d \tilde{B}_R^c - y_\tau A_\tau \tilde{L} \cdot H_d \tilde{L}_R^c + h.c.) .
\end{eqnarray}


The NMSSM superpotential possesses a global $Z_3$ symmetry which is broken during the electroweak phase transition in the early universe.  The resulting domain walls should disappear before nucleosynthesis.  However $Z_3$ breaking (via singlet tadpoles) leads to a vacuum expectation value (vev) for the singlet that is much larger than the electroweak scale.  Thus the requirement of the fast disappearance of the domain walls appears to destabilize the hierarchy of vevs in the NMSSM.  Fortunately, in Ref.s ~ \cite{Panagiotakopoulos:1998yw, Panagiotakopoulos:2000wp} is was shown that, by imposing a $Z_2$ R-symmetry, both the domain wall and the stability problems can be eliminated.  Following \cite{Panagiotakopoulos:1998yw}, we assume that tadpoles are induced, but they are small and their effect on the phenomenology is negligible.


We use supergravity motivated boundary conditions to parametrize the soft masses and tri-linear couplings.  Defining a constrained version of the NMSSM, we assume unification of the gaugino masses
\begin{eqnarray}
 M_1 = M_2 = M_3 = M_{1/2} ,
\end{eqnarray}
the sfermion and Higgs masses (but not the singlet mass)
\begin{eqnarray} 
 M_{Q}^2 = M_{T_R}^2 = M_{B_R}^2 = M_{L}^2 = M_{L_R}^2 = M_{H_u}^2 = M_{H_d}^2 = M_0^2 ,
\end{eqnarray}
and the tri-linear couplings 
\begin{eqnarray}  
 A_t = A_b = A_\tau = A_\kappa = A_\lambda = A_0 , 
\end{eqnarray}
at the scale of a grand unified theory (GUT) where the three standard gauge couplings meet $g_1 = g_2 = g_3 = g_{GUT}$.  This leaves six parameters in the model: $M_0$, $M_{1/2}$, $A_0$, $M_S$, $\lambda$ and $\kappa$.  Electroweak symmetry breaking introduces the Higgs and singlet vevs, $\langle H_u \rangle$, $\langle H_d \rangle$, $\langle S \rangle$. From Eq.\ref{eq:W_NMSSM} we see that when the singlet acquires a vev, the MSSM $\mu$ term is dynamically generated as $\mu = \lambda \langle S \rangle$, and thus the NMSSM naturally solves the $\mu$ problem.
The three minimization equations for the Higgs potential \cite{Hugonie:2007vd} and 
\begin{eqnarray}  
 \langle H_u \rangle^2 + \langle H_d \rangle^2 = v^2 
\end{eqnarray}
(here $v = \sqrt{2} /g_2$ is the standard Higgs vev) eliminate four parameters.  Thus our constrained NMSSM model has only five free parameters and a sign.  Defining $\tan\beta = \langle H_u \rangle/\langle H_d \rangle$, the parameters of the next-to-minimal supergravity motivated model (NmSuGra) are
\begin{eqnarray}
 P = \{M_0, M_{1/2}, A_0, \tan\beta, \lambda, {\rm sign}(\mu)\} .
\label{eq:5Para}
\end{eqnarray}

 
Constrained versions of the NMSSM have been studied in the recent literature. 
The most constrained version is the cNMSSM \cite{Djouadi:2008yj} with $M_S = 
M_0$.  In other cases the $A_\kappa = A_\lambda$ relation is relaxed 
\cite{Hugonie:2007vd}, and/or $\kappa$ is taken as a free parameter 
\cite{Belanger:2005kh,Cerdeno:2007sn}, or the soft Higgs masses are allowed to 
deviate from $M_0$ \cite{Djouadi:2008uw} giving less constrained models.  In the 
spirit of the CMSSM/mSuGra, we adhere to universality and use only $\lambda$ to 
parametrize the singlet sector.  This way, we keep all the attractive features of 
the CMSSM/mSuGra while the minimal extension alleviates problems rooted in the 
MSSM, making the NMSSM a more natural model.

As we have shown in our previous work \cite{Balazs:2008ph}, NmSuGra phenomenology bears a high similarity to the minimal supergravity motivated model.  The most significant departures from a typical mSuGra model are the possibility of a singlino-dominated neutralino and the extended Higgs sector, which may provide new resonance annihilation channels and Higgs decay channels, potentially weakening the mass limit from LEP.  The majority of NmSuGra phenomenology can be described in terms of the familiar mSuGra features \cite{Baer:2003yh}.  The gross structure of the parameter space, for example, can be easily understood in terms of the predominant neutralino (co-)annihilation mechanisms.  NmSuGra features the well known neutralino-slepton co-annihilation, Higgs resonance corridor annihilation and focus point regions, as well as small regions of mostly singlino-type neutralino annihilation via multiple channels.  In the following discussions, we will rely on the similarity between our model and mSuGra to interpret the results of the likelihood scan, while paying special attention to all distinct phenomenological signatures.

\section{Bayesian inference}
\label{sec:Bayes}

Since several excellent papers have appeared on this subject recently \cite{Feroz:2008wr, Trotta:2008bp, Cabrera:2008tj, Feroz:2009dv, AbdusSalam:2009qd}, in this section, we summarize the concepts of Bayesian inference that we use in our analysis in a compact fashion.  Our starting hypothesis $H$ is the validity of the NmSuGra model.  The conditional probability $\mathcal{P}(P|D;H)$ quantifies the validity of our hypothesis by giving the chance that the NmSuGra model reproduces the available experimental data $D$ with its parameters set to values $P$.  When this probability density is integrated over a region of the parameter space it yields the posterior probability that the parameter values fall into the given region.


Bayes' theorem provides us with a simple way to calculate the posterior probability distribution as
\begin{eqnarray}
 \mathcal{P}(P|D;H) = \mathcal{P}(D|P;H) \frac{\mathcal{P}(P|H)}{\mathcal{P}(D|H)} .
\end{eqnarray}
Here $\mathcal{P}(D|P;H)$ is the likelihood that the data is predicted by NmSuGra with a specified set of parameters.  The a-priori distribution of the parameters within the theory $\mathcal{P}(P|H)$ is fixed by purely theoretical considerations independently from the data.  The evidence $\mathcal{P}(D|H)$ gives the probability of the hypothesis in terms of the data alone.  The latter can easily be seen from Bayes' theorem by multiplying both sides with $\mathcal{P}(D|H)$ and integrating for the full parameter space:
\begin{eqnarray}
 \mathcal{P}(D|H) = \int \mathcal{P}(D|P;H) \mathcal{P}(P|H) dP .
\end{eqnarray}

If the data under consideration are statistically independent, as in our case,
the likelihood function factorizes
\begin{eqnarray}
 \mathcal{P}(D|P;H) = \prod_i {\cal L}_i(D,P;H) ,
\end{eqnarray}
where $\mathcal{L}_i$ is the likelihood associated with the $i$th measurable, and is given as a convolution
\begin{eqnarray}
 {\cal L}_i(D,P;H) = {\cal L}_i^E(D) \otimes {\cal L}_i^T (P;H) .
\end{eqnarray}
If the experimental and theoretical likelihoods corresponding to the $i$th measurable, ${\cal L}_i^E$ and ${\cal L}_i^T$, are normally distributed, then the likelihood function is a Gaussian
\begin{eqnarray}
 {\cal L}_i(D,P;H) = \frac{1}{\sqrt{2 \pi} \sigma_i} \exp(\chi_i^2(D,P;H)/2) .
\end{eqnarray}
In this case the exponents
\begin{eqnarray}
 \label{eq:chi2}
 \chi_i^2(D,P;H)/2 = (d_i - t_i(P;H))^2/2 \sigma_i^2 ,
\end{eqnarray}
are defined in terms of the experimental data $D = \{d_i \pm \sigma_{i,e}\}$ and theoretical predictions $T = \{t_i \pm \sigma_{i,t}\}$ for these measurables.  Independent experimental and theoretical uncertainties combine into $\sigma_i^2 = \sigma_{i,e}^2 + \sigma_{i,t}^2$.
In cases when the experimental data only specify a lower (or upper) limit, the corresponding likelihood function can be written in terms of the error function
\begin{eqnarray}
 \mathcal{L}_i(D|P;H) = \frac{1}{2} {\rm erfc} (\pm \sqrt{\chi_i^2(D,P;H)/2}) .
\end{eqnarray}

Despite its dependence on several parameters, the likelihood function can easily be visualized by plotting profile likelihood distributions.  These are constructed by finding the maximum likelihood hypersurface 
\begin{eqnarray}
 \mathcal{L}^{max}(D|p_i;H) = \max_{p_1,...,p_{i-1},p_{i+1},...,p_n}(\mathcal{L}(D|P;H)) .
\end{eqnarray}
and projecting this to one (or more functions) of the parameters.  These profile likelihoods highlight the model regions where the likelihood is highest and lowest.  Although this is certainly of interest, in the Bayesian context the estimated values of the parameters are determined by the posterior probability density.


While the posterior probability density $\mathcal{P}(P|D;H)$ depends on all the parameters $P = \{p_i\}$ of NmSuGra, it is useful to know the probability distribution of each single parameter $p_i$.  This latter is referred to as the marginalized probability and given by 
\begin{eqnarray}
 \mathcal{P}(p_i|D;H) = \int \mathcal{P}(P|D;H) dp_1...dp_{i-1}dp_{i+1}...dp_n .
 \label{eq:margin1}
\end{eqnarray}
Similarly to this, marginalization can be performed to two (or more) variables resulting in a two (or more) dimensional distribution:
\begin{eqnarray}
 \mathcal{P}({p_i,p_j}|D;H) = \int \mathcal{P}(P|D;H) dp_1...dp_{i-1}dp_{i+1}...dp_{j-1}dp_{j+1}...dp_n .
 \label{eq:margin2}
\end{eqnarray}
Our main goal in this work is to evaluate marginalized probability distributions for the five theoretical parameters of the NmSuGra model.

Marginalization can also be performed to an arbitrary function (or several functions) of the parameters.  The posterior probability density of an arbitrary function (of a subset) of the parameters $f(P)$ is obtained as 
\begin{eqnarray}
 \mathcal{P}(f|D;H) = \int \delta(f-f(P)) \mathcal{P}(P|D;H) dP .
\end{eqnarray}
These marginalized probability distributions are useful when we compare various NmSuGra predictions to the experimental likelihood distributions to check the consistency of the model with the data, and when we assess the future detectability of the model.

It is also useful to introduce confidence level regions measured by relative probabilities.  We define an $x$ percent confidence level region ${\mathcal R}_x$ by the set of the minimal parameter regions supporting $x$ percent of the total probability:
\begin{eqnarray}
 x = {\left(\int_{{\mathcal R}_x} \mathcal{P} dP\right)}{\left(\int \mathcal{P} dP\right)^{-1}} .
\end{eqnarray}
Here $\mathcal{P}$ can be a likelihood function or a posterior probability distribution, and the integral in the denominator extends to the full parameter space.


Profile likelihood distributions and posterior probability distributions carry different information about the theoretical parameter space.  While the former expresses in which regions the model can or cannot fit the data, the latter gives the probability of the of a given parameter region in the light of the data.  As Bayes theorem shows these two differ by the factors of the theoretical prior and the evidence.  
Setting the evidence aside as a trivial normalization factor, the prior should come from purely theoretical considerations, from an underlying theory perhaps.  In the case of NmSuGra this information is highly uncertain and the a prior can be selected, at best, based on simplistic assumptions such as fine-tuning.  For this reason in this work we resort to a trivial (uniform) prior, expressing the fact that we have no reliable theoretical information about the a-priori probability distribution over the parameter space.

Besides this difference marginalized posterior probabilities also capture information about the size and structure of the parameter space via the integration in Eq.s (\ref{eq:margin1}) and (\ref{eq:margin2}).  In other words the volume of the parameter space, via the integration measure, affects the posterior probability.  The likelihood profile do not contain this information.  Thus, even with a prior uniform over the parameter space, this volume effect can substantially change the shape of posterior probability distributions compared to the profile likelihoods.

\section{Likelihood analysis of NmSuGra}
\label{sec:likelihood}

Our main intent is to calculate the posterior probability distributions for the five continuous parameters of NmSuGra and check the consistency of the model against available experimental data.  To this end, we use the publicly available computer code NMSPEC \cite{Ellwanger:2006rn} to calculate the spectrum of the superpartner masses and their physical couplings from the model parameters given in Eq.~(\ref{eq:5Para}).  Then, we use NMSSMTools 2.1.0 and micrOMEGAs 2.2 \cite{Belanger:2006is} to calculate the abundance of neutralinos ($\Omega h^2$), the spin-independent neutralino-proton elastic scattering cross section ($\sigma_{SI}$), the NmSuGra contribution to the anomalous magnetic moment of the muon ($\Delta a_{\mu}$), and various b-physics related quantities.  All the experimental data $D$ used in our likelihood analysis are listed in Table \ref{tab:Data}.  Uncertainties arising in the supersymmetric calculation of $\Delta a_{\mu}$ and the b-physics related quantities are calculated using NMSSMTools.  Among the standard input parameters, $m_b(m_b) = 4.214$ GeV and $m_t^{pole}=171.4$ GeV are used.

As alluded to previously, the LEP mass limit for the standard model Higgs $m_h>114.4$ GeV is not applicable to the lightest scalar in the NMSSM.  Specifically, any component of the gauge-singlet scalar in the lightest Higgs would decrease its couplings to gauge bosons.  We calculate the modified limit according to Ref. \cite{Barger:2006dh}, by comparing the NMSSM coupling of the lightest scalar to the $Z$ boson with that of the standard model.  Furthermore, in the NMSSM there is the possibility of a light pseudoscalar field $a$ with $m_a<m_W$ to which the lightest Higgs could decay as $h \to aa$ without subsequent charged decays.  For these rare occurances the LEP Higgs limit is further relaxed.

\TABULAR[thb]{llrr}
{ Observable & Limit type & $d_i \pm \sigma_{i,e}$ & $\sigma_{i,t}^{SUSY}$ \\
\hline
                       $m_h$ &   lower limit &                                up to 114.4 GeV   \cite{Barger:2006dh} &    3.0 GeV       \cite{Frank:2006yh} \\
        $m_{\tilde{\tau}_1}$ &   lower limit &                         73.0 or$^1$ 87.0 GeV \cite{AbdusSalam:2009qd} &      10 \%                           \\
           $m_{\tilde{e}_R}$ &   lower limit &                         73.0 or$^1$ 100. GeV \cite{AbdusSalam:2009qd} &      10 \%                           \\
         $m_{\tilde{\mu}_R}$ &   lower limit &                         73.0 or$^1$ 95.0 GeV \cite{AbdusSalam:2009qd} &      10 \%                           \\
         $m_{\tilde{\nu}_e}$ &   lower limit &                         43.0 or$^1$ 94.0 GeV \cite{AbdusSalam:2009qd} &      10 \%                           \\
           $m_{\tilde{t}_1}$ &   lower limit &                         65.0 or$^1$ 95.0 GeV \cite{AbdusSalam:2009qd} &      10 \%                           \\
           $m_{\tilde{b}_1}$ &   lower limit &                         59.0 or\footnote{If $m_{sfermion}-m_{Z_1}\lessgtr 10$ GeV the ${}^{lower}_{higher}$ value applies.} 95.0 GeV \cite{AbdusSalam:2009qd} &      10 \%                           \\
           $m_{\tilde{q}_1}$ &   lower limit &                                    318.0 GeV \cite{AbdusSalam:2009qd} &      10 \%                           \\
           $m_{\tilde{W}_1}$ &   lower limit &                         43.0 or\footnote{If $m_{\tau_\nu}\lessgtr 300$ GeV the ${}^{lower}_{higher}$ value applies.} 92.4 GeV \cite{AbdusSalam:2009qd} &      10 \%                           \\
           $m_{\tilde{Z}_1}$ &   lower limit &                                     50.0 GeV \cite{AbdusSalam:2009qd} &      10 \%                           \\
             $m_{\tilde{g}}$ &   lower limit &                                    195.0 GeV \cite{AbdusSalam:2009qd} &      10 \%                           \\
            $\Delta a_{\mu}$ & central value &           (29.0 $\pm$ 9.0)$\times 10^{-10}$ \cite{Jegerlehner:2009ry} & negligible                           \\
                $\Delta m_d$ & central value & (5.07 $\pm$ 0.04)$\times 10^{11}$ ps$^{-1}$    \cite{Barberio:2008fa} &       1 \%    \cite{Belanger:2006is} \\ 
         $B(b \to s \gamma)$ & central value &           (3.50 $\pm$ 0.17)$\times 10^{-4}$    \cite{Barberio:2008fa} &      10 \%      \cite{Misiak:2006zs} \\
 $B(B^+ \to \tau+ \nu_\tau)$ & central value &            $(1.73 \pm 0.35) \times 10^{-4}$      \cite{Artuso:2009jw} &      10 \%    \cite{Belanger:2006is} \\ 
     $B(B_s \to \mu^+\mu^-)$ &   upper limit &                        4.7 $\times 10^{-8}$      \cite{Artuso:2009jw} &      10 \%       \cite{Buras:2002vd} \\
                $\Omega h^2$ &   upper limit &                         0.1143 $\pm$ 0.0034     \cite{Komatsu:2008hk} &      10 \%    \cite{Belanger:2006is} \\
               $\sigma_{SI}$ &   upper limit &                                   CDMS 2008       \cite{Ahmed:2008eu} &      20 \%       \cite{Ellis:2008hf} \\
\label{tab:Data}}
{Observables used in the calculation of the posterior probability distribution $\mathcal{P}(P|D;H)$.  Experimental data are listed under column $d_i \pm \sigma_{i,e}$, and typical uncertainties related to the NmSuGra calculations are under $\sigma_{i,t}^{SUSY}$.}

Using the above specified tools, we generate theoretical predictions for NmSuGra in the following part of its parameter space:
\begin{eqnarray}
 0 < M_0 < 5 ~{\rm TeV}, ~~~ 0 < M_{1/2} < 2 ~{\rm TeV}, ~~~ -3 ~{\rm TeV} < A_0 < 5 ~{\rm TeV}, \nonumber \\ 
 0 < \tan\beta < 60, ~~~ 10^{-5} < \lambda < 0.6, ~~~ {\rm sign}(\mu) > 0. ~~~~~~~~~~~~
\end{eqnarray}
In this work, we only consider the positive sign of $\mu$ because, similarly to mSuGra \cite{Feroz:2008wr}, the likelihood function is suppressed by $\Delta a_\mu$ and $B(b \to s \gamma)$ in the negative $\mu$ region.
When calculating posterior probabilities, we use a uniform prior; or equivalently, we define the measure of the integration in Eq.s (\ref{eq:margin1}) and (\ref{eq:margin2}) by weighting all parts of the parameter space equally.  We evaluate the integrals over the likelihood function in two different ways.  In the first method, we simply select random model points from the parameter space according to a uniform distribution.  In the second, we adopt the Markov Chain Monte Carlo technique as described in \cite{Baltz:2006fm}.  The Markov Chain implements the Metropolis algorithm, and is more efficient than the random scan in terms of the proportion of viable points scanned.  However, by construction the Markov Chain has a tendency to focus strongly on high-likelihood regions and so the results of individual chains can be somewhat sporadic.  As the five dimensional NmSuGra parameter space is not prohibitively large, the uniform random scan gives a good means of checking the Markov Chain as it provides very consistent results over different runs, although with somewhat less fine detail in the likelihood distribution.  In general, for all of our results the two different methods are in good agreement.

\subsection{Profile likelihoods}

Turning to our numerical results, in Figure \ref{fig:PLD.inputs} we show profile likelihood distributions projected to individual input parameters.  In these plots, for ease of interpretation, we normalize the profile likelihoods such that a model point fitting perfectly the data (with $\chi^2 = 0$) would result in a likelihood value of 1. 

These likelihood profiles tell us the best fit of NmSuGra predictions to experimental and where in the parameter space they occur.  Regions of high likelihood are not difficult to find within NmSuGra.  For our results, the model point with the highest likelihood has a $\chi^2_{total} = 1.85$ and there exist ample parameter regions with the most likely $\chi^2/d.o.f \approx 1$.

\FIGURE[th]{
\epsfig{file=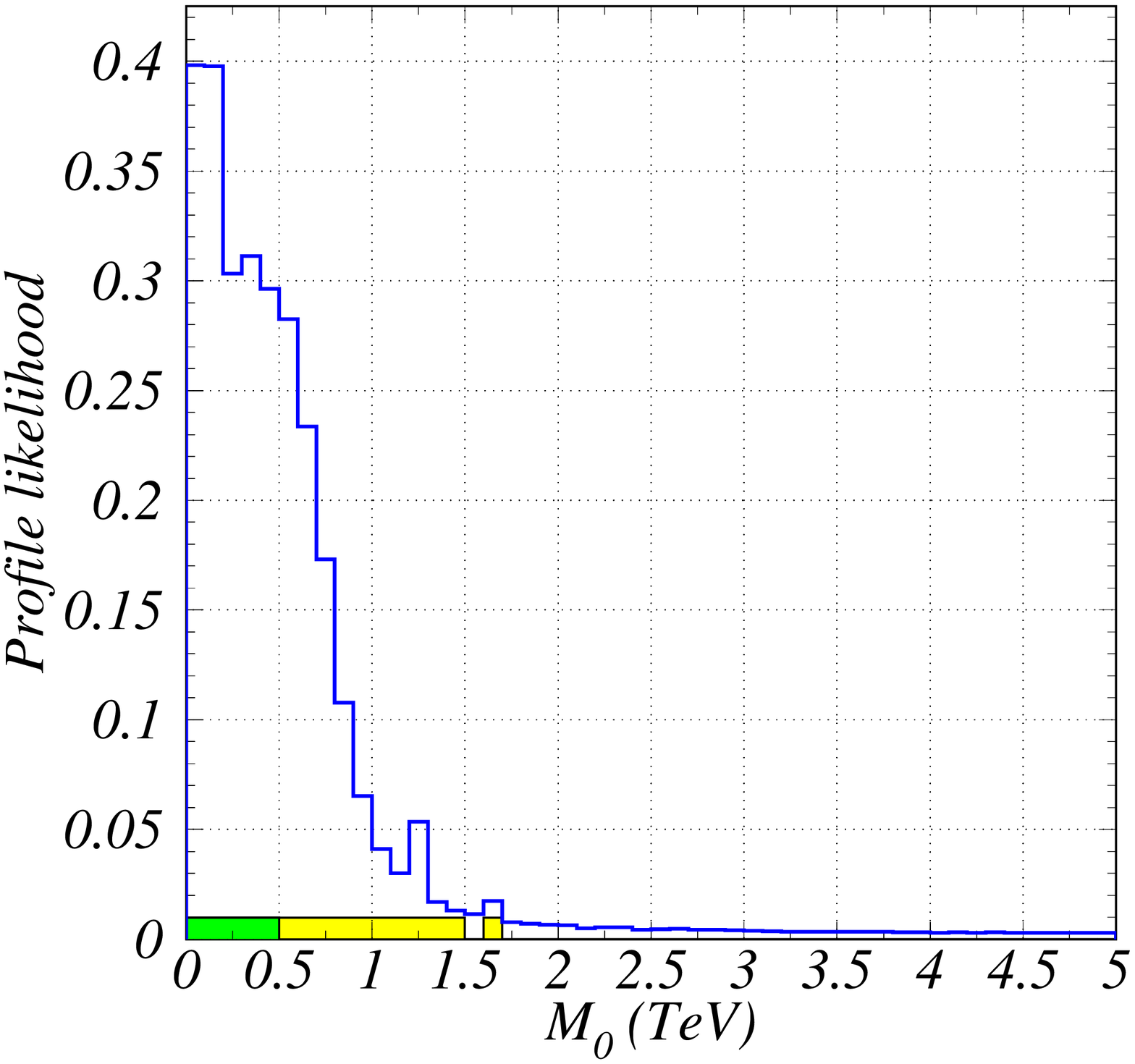,width=0.49\textwidth}
\epsfig{file=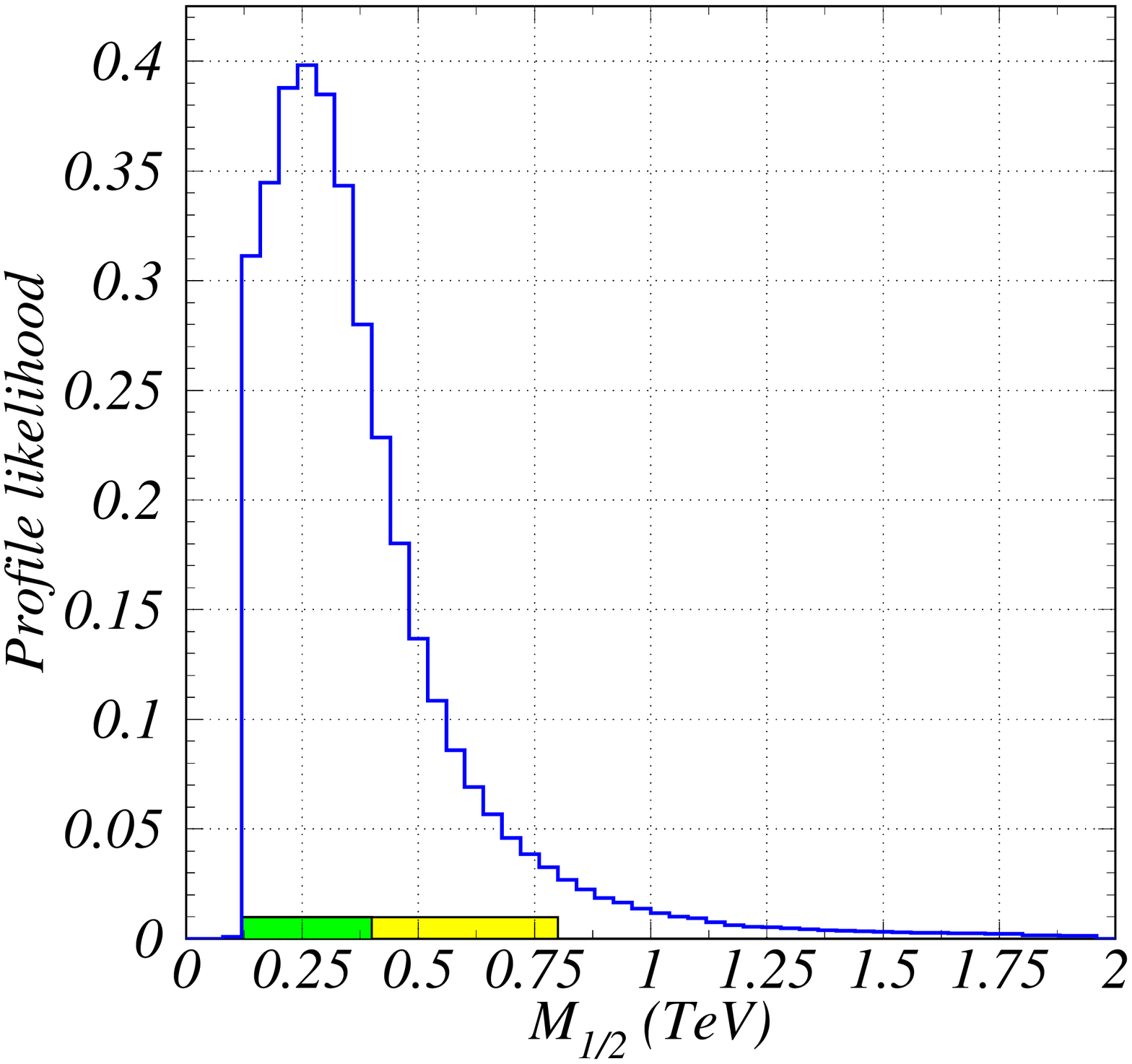,width=0.49\textwidth}
\epsfig{file=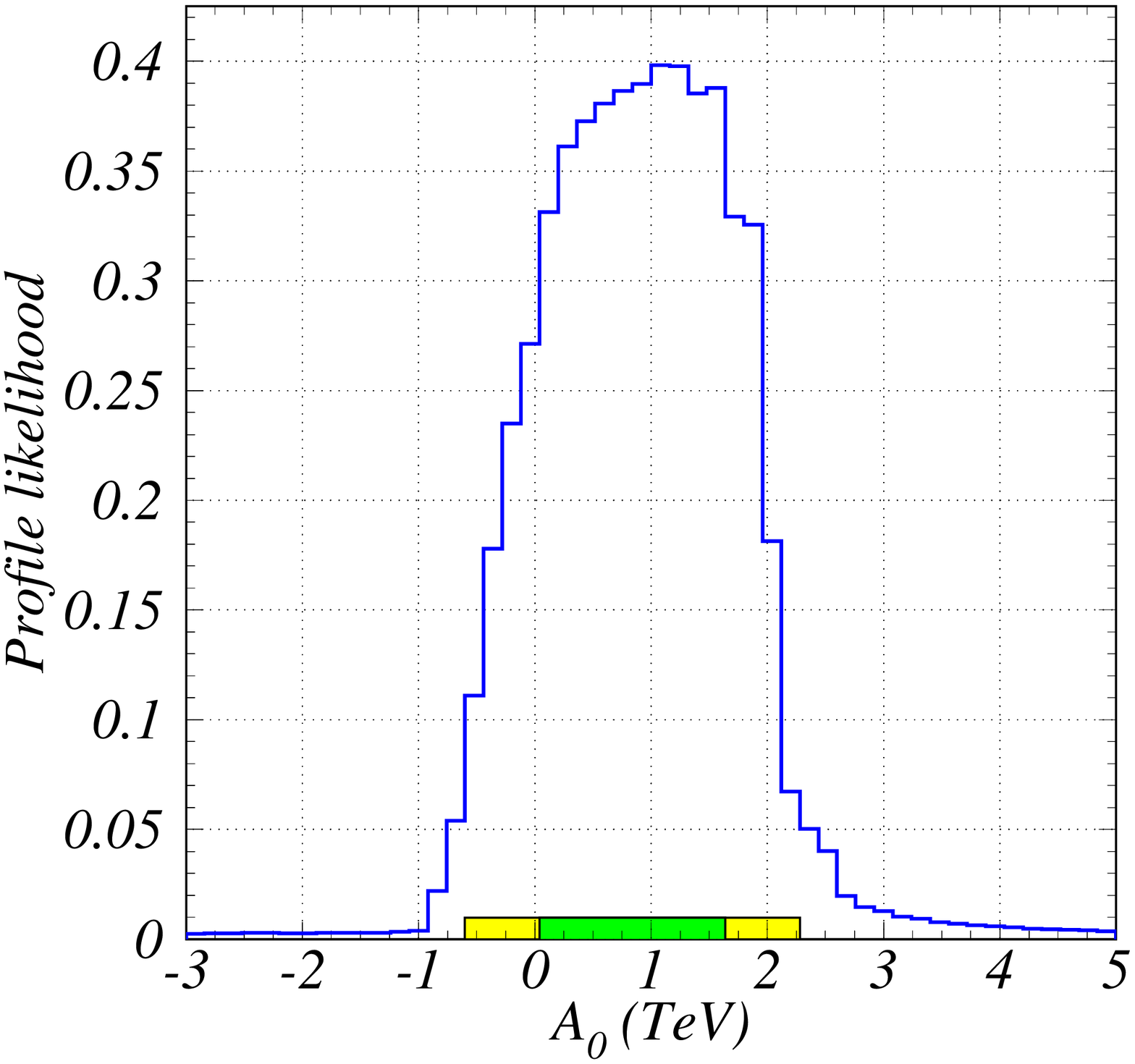,width=0.49\textwidth}
\epsfig{file=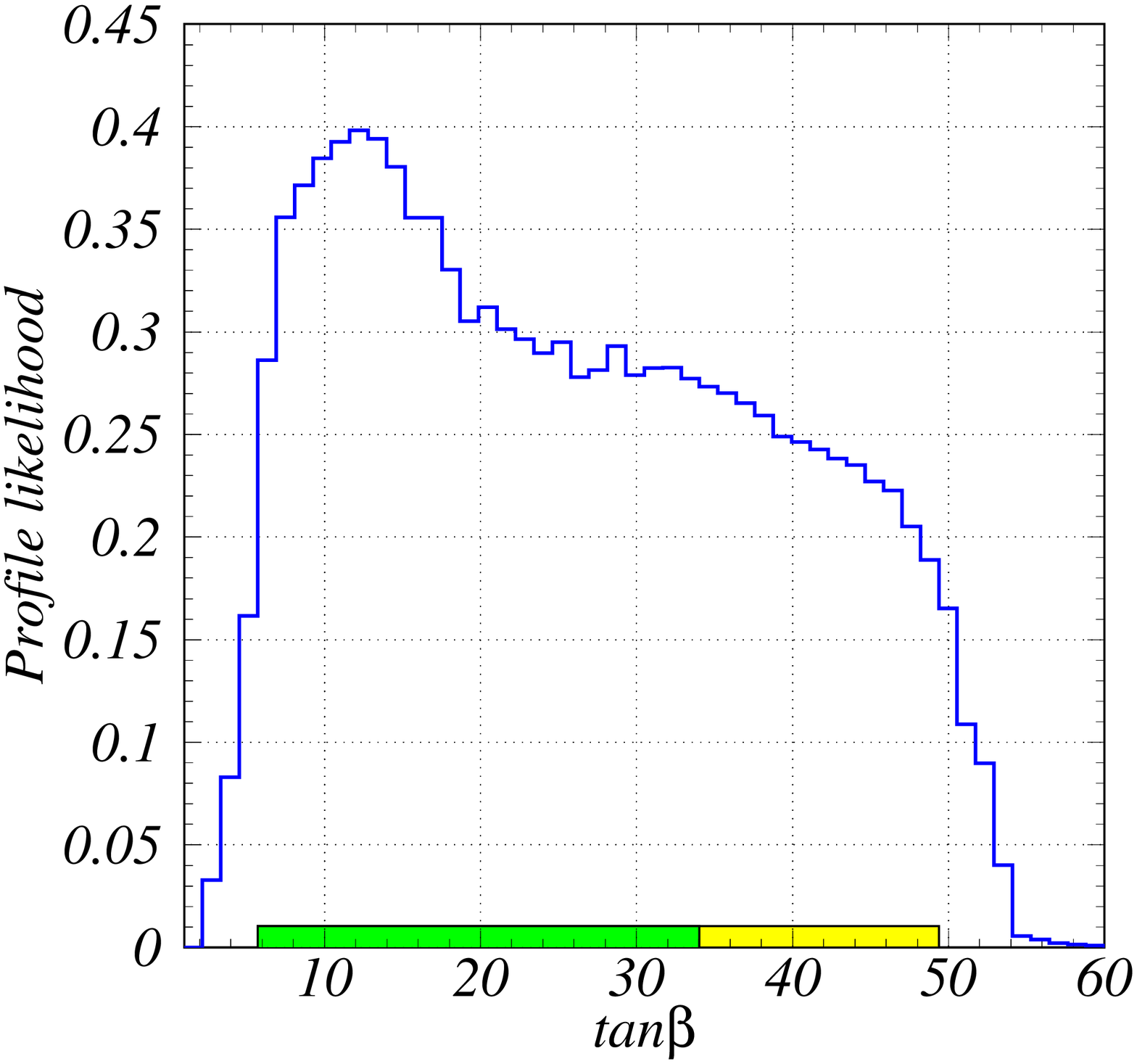,width=0.49\textwidth}
\caption{Profile likelihood distributions of the NmSuGra input parameters.  Green (yellow) coloring indicates 68 (95) percent confidence level regions.}
\label{fig:PLD.inputs}}

The top left frame of Figure \ref{fig:PLD.inputs} indicates that the highest likelihood regions lie at low values of the common scalar mass $M_0$, within $0 < M_0 \leq 0.5$ TeV ($0 < M_0 \leq 1.5$ TeV) at 68 (95) \% confidence level.  This happens because, with the exception of some of the b-physics observables, the experimental data mostly favor the sfermion-neutralino co-annihilation region, lying around low $M_0$ values.  Higgs resonance corridors at intermediate $M_0$ still give a reasonable fit, while the likelihood function tends to drop in the focus point region at high $M_0$ values due mainly to $\Delta a_\mu$ disfavoring high sparticle masses (consistent with our earlier results \cite{Balazs:2008ph}).  

Although the highest likelihood is about $3.5 \times 10^{-3}$ at $M_0 = 5$ TeV, this corresponds to a combined $\chi^2_{total} < 12$.  Discarding direct (s)particle mass limits (which don't affect the focus point) and counting only the last seven entries in Table \ref{tab:Data}, this corresponds to $\chi^2/d.o.f < 2$ in average, which is not a bad fit.  This means that viable model points remain in the focus point, although with less likelihood than those at lower $M_0$ values.

We can draw a similar conclusion for the common gaugino mass $M_{1/2}$ from the next frame of Figure \ref{fig:PLD.inputs}.  The data appear to prefer small to moderate values of $M_{1/2}$ for NmSuGra falling in the region $0.18 \leq M_{1/2} \lesssim 0.35$ TeV ($0.18 \leq M_{1/2} \lesssim 0.8$ TeV) with a 68 (95) \% confidence level.  The likelihood is suppressed at high $M_{1/2}$ by $\Delta a_\mu$.  

The same is evident from the profile likelihood of the common tri-linear parameter $A_0$, which has two maximal regions, one slightly negative but close to zero and another around 1 TeV.  The data show a high preference toward $\tan\beta \sim 5$ and somewhat lower toward the $5 \lesssim \tan\beta \lesssim 35$ region.  The Higgs-singlet-Higgs coupling $\lambda$ (not shown) has a rather featureless likelihood profile tapering down above $\lambda > 0.5$ at a 95 \% confidence level.

\FIGURE[th]{
\epsfig{file=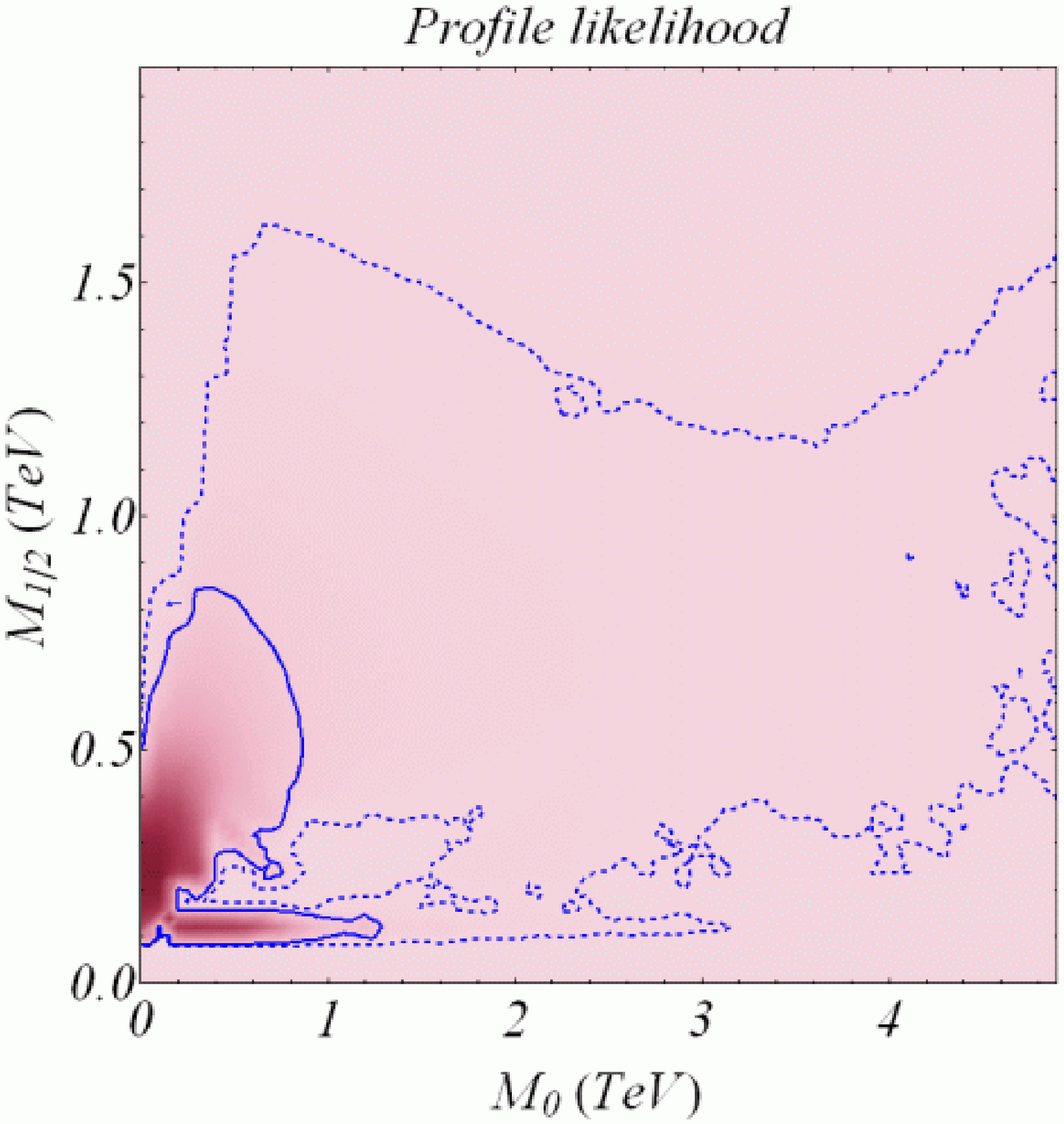,width=0.49\textwidth}
\epsfig{file=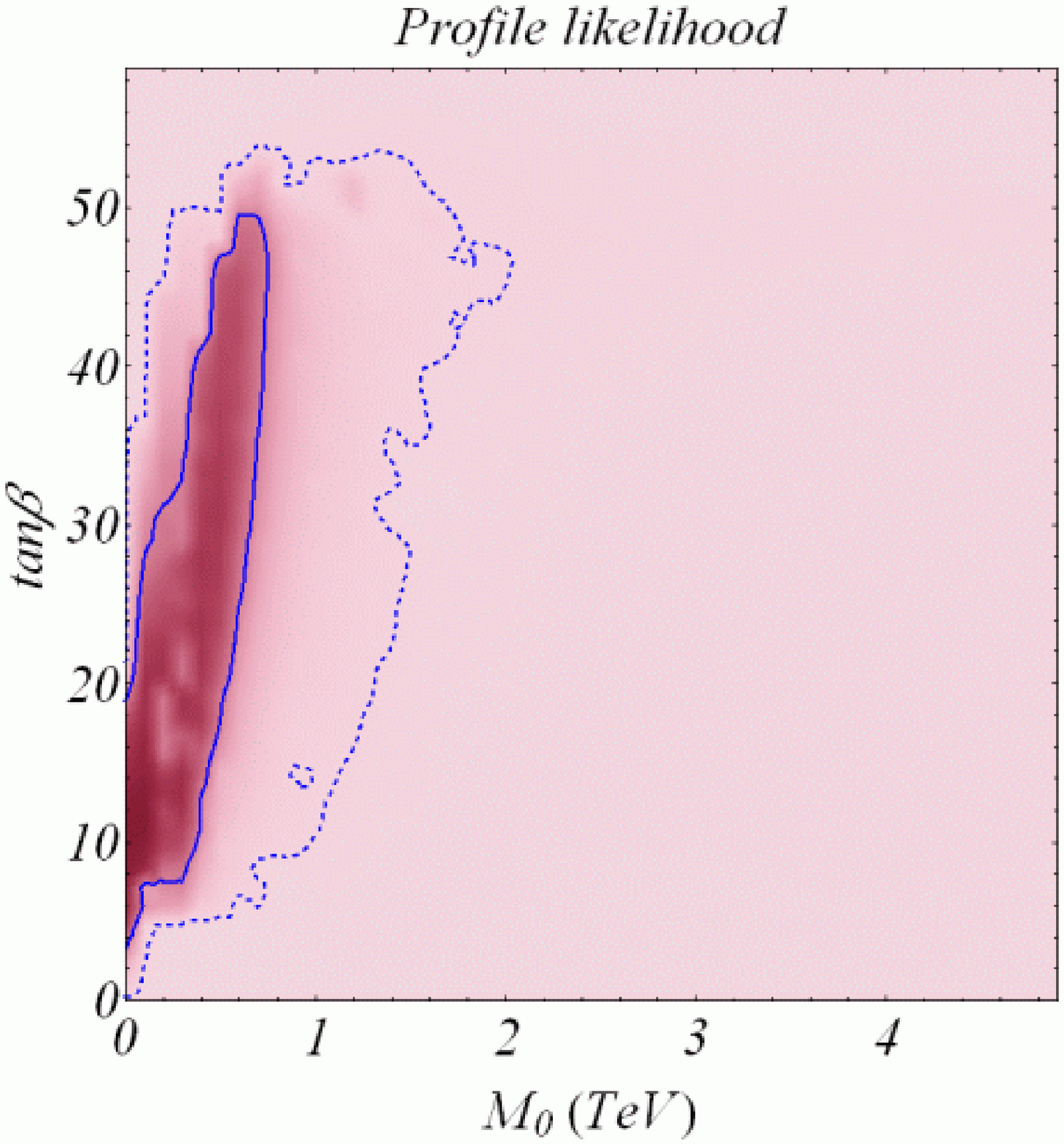,width=0.49\textwidth}
\epsfig{file=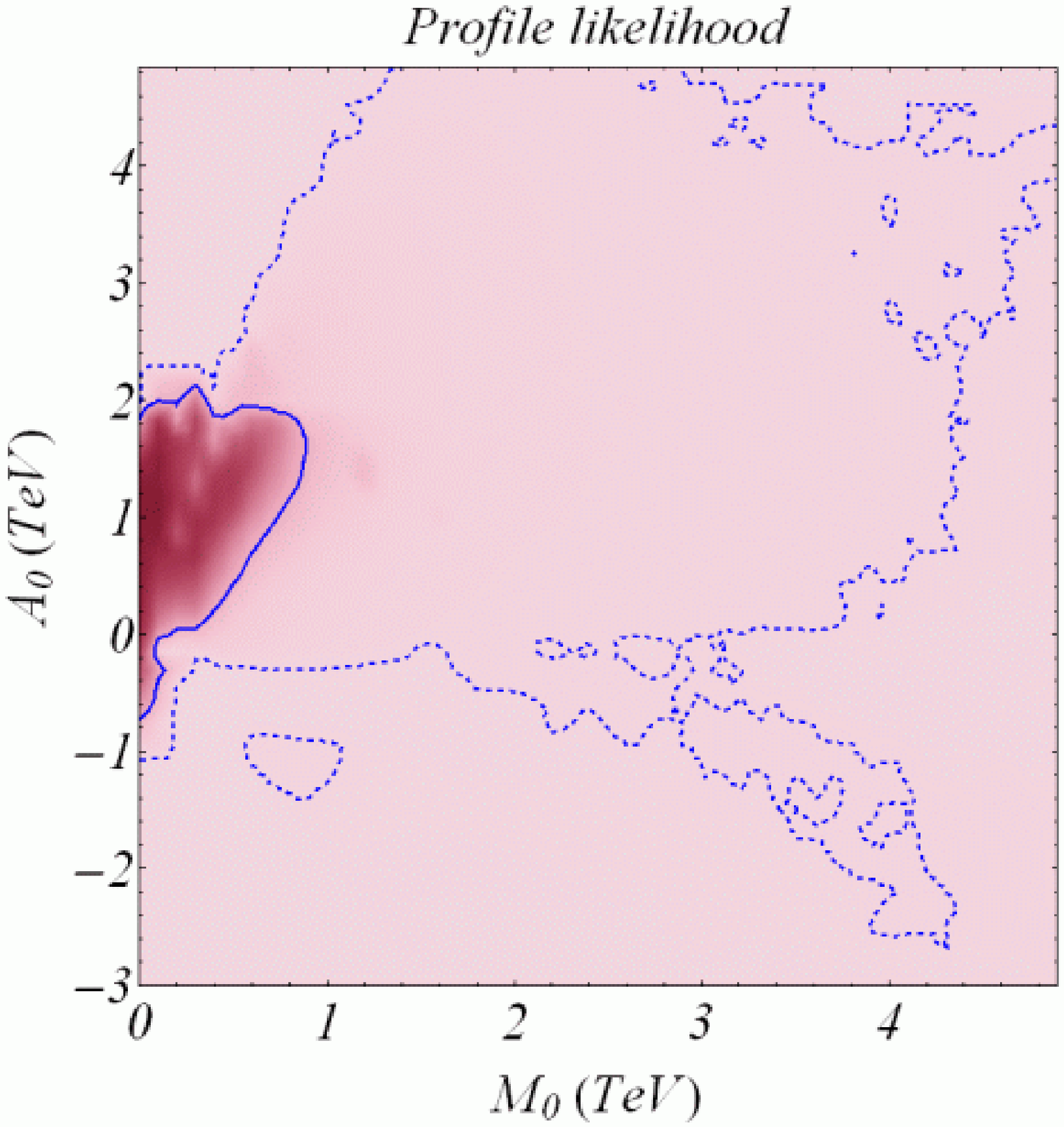,width=0.49\textwidth}
\epsfig{file=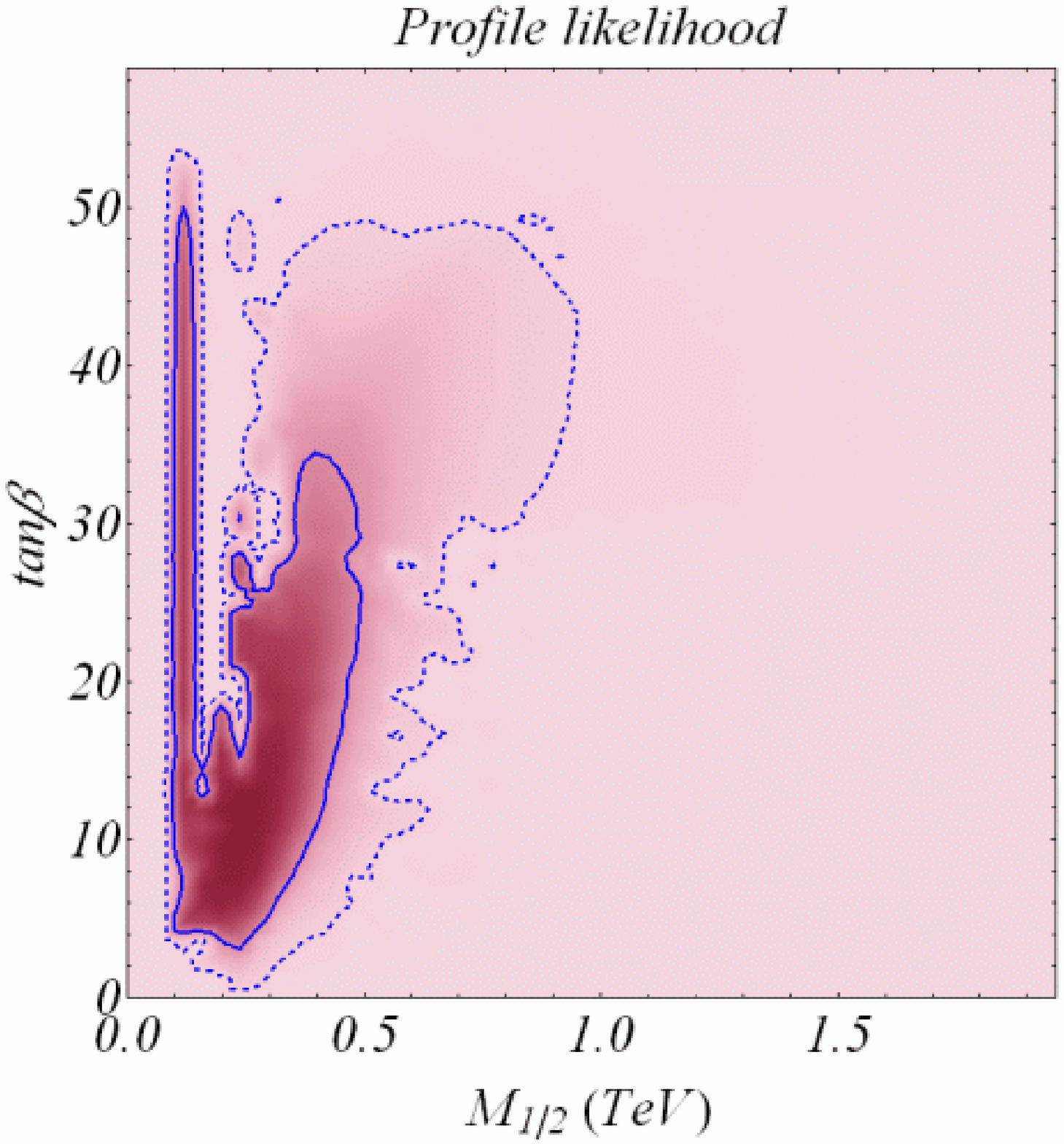,width=0.49\textwidth}
\caption{Profile likelihood distributions as the function of pairs of NmSuGra input parameters.  The higher likelihood regions are darker.  Solid (dotted) blue lines indicate 68 (95) percent confidence level contours.}
\label{fig:PLD.inputs2d}}

In Figure \ref{fig:PLD.inputs2d} we show profile likelihood distributions as the function of various pairs of NmSuGra input parameters.  Solid (dotted) blue lines indicate 68 (95) percent confidence level contours.  According to the first frame the highest likelihoods coincide with the stau-neutralino and the adjoint (pseudoscalar and heavier) Higgs resonance region.  Annihilation via the lightest Higgs in the s-channel is also prominent.  The last frame supports this giving us the insight that stau-neutralino co-annihilation is dominant at low to moderate $\tan\beta$, while resonant annihilation via the lightest Higgs boson is essentially independent of $\tan\beta$.

The first three frames clearly indicate that the likelihood in the focus point, for high $M_0$, is suppressed by the anomalous magnetic moment of the muon, as we found earlier.  The relatively featureless second frame is included for reference and completness; later we will use the posterior probability conterpart of this distribution to confirm the location of the focus point at high $\tan\beta$.  Although the asymmetry is moderate, most of the high likelihood region has positive $A_0$ as can be read from the second frame.

\subsection{Posterior probabilities}

While the likelihood function may peak in certain regions of the parameter space, signaling that the theory predictions fit well the experiment, this alone does not give information on the probability of various parameter regions.  It may happen, for example, that the theory has high (moderate) likelihood over a small (large) parameter region.  According to Bayes theorem the probability assigned to a parameter region is the accumulated likelihood over the given region of parameter space.

Furthermore, if a theory has many parameters, or more generally a large parameter space, then one could expect a better fit to experiment would be possible compared to the case with a fewer theoretical parameters.  However such a model is arguably less natural (or possibly more contrived).  A marginalized posterior probability depends on the size of the parameter space and the cumulative magnitude of the likelihood function over this space, rather than simply peak values.  As a consequence, larger parameter spaces with low likelihoods incur an inherent penalty to the posterior probability as unitarity suppresses the posterior probability, in accordance with Occam's principle.

We show the posterior probability marginalized to the five input parameters in Figure \ref{fig:MPP.inputs}.  Since the integral of the posterior probability distribution marginalized to a given parameter $p_i$ 
\begin{eqnarray}
 \mathcal{P}(a < p_i < b|D;H) = \int_a^b \mathcal{P}(p_i|D;H) dp_i ,
\end{eqnarray}
yields the probability that $p_i$ falls in the interval $[a,b]$, we normalize the marginalized posterior densities such that the area under the distribution is unity.

\FIGURE[th]{
\epsfig{file=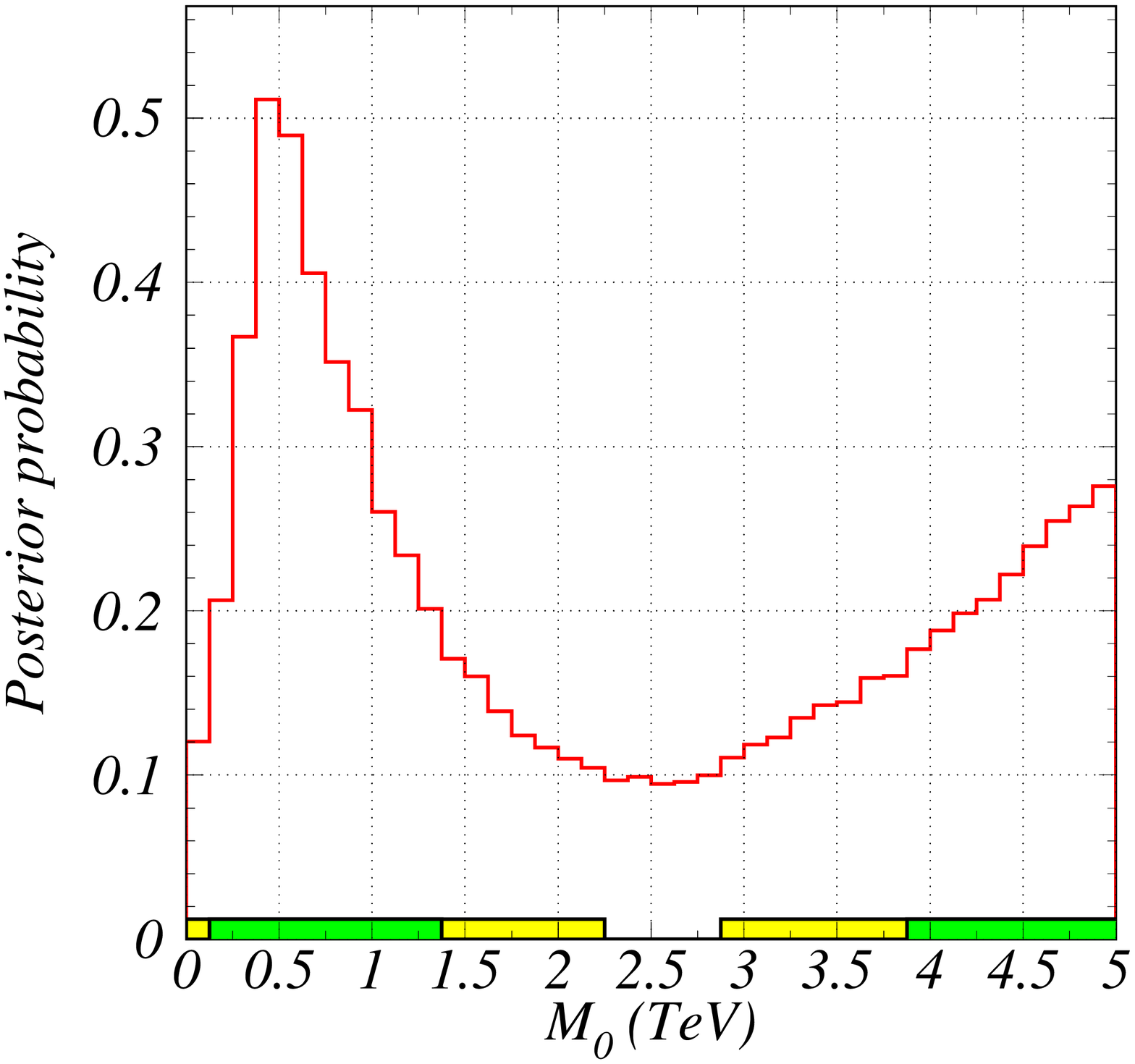,width=0.49\textwidth}
\epsfig{file=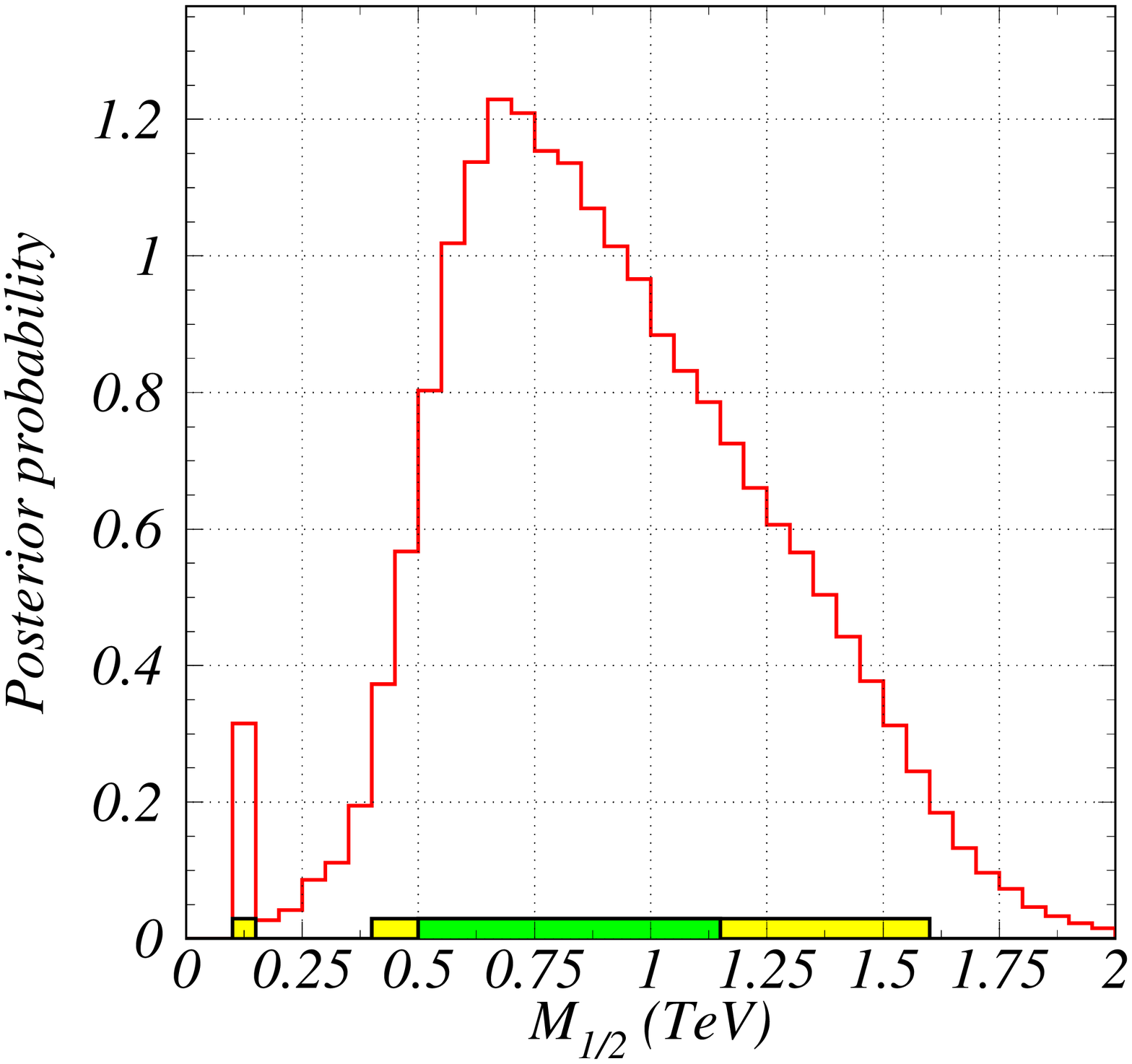,width=0.49\textwidth}
\epsfig{file=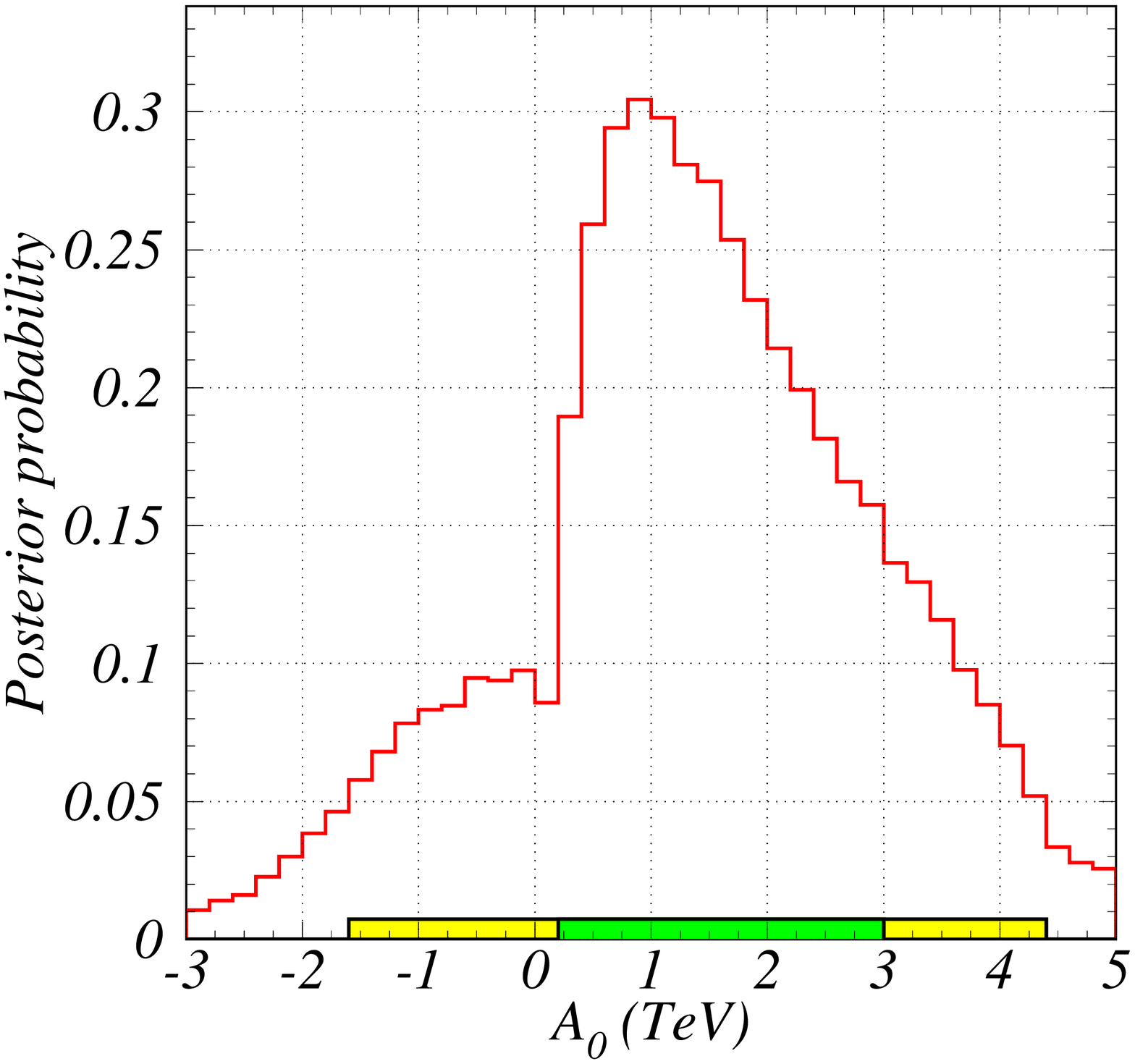,width=0.49\textwidth}
\epsfig{file=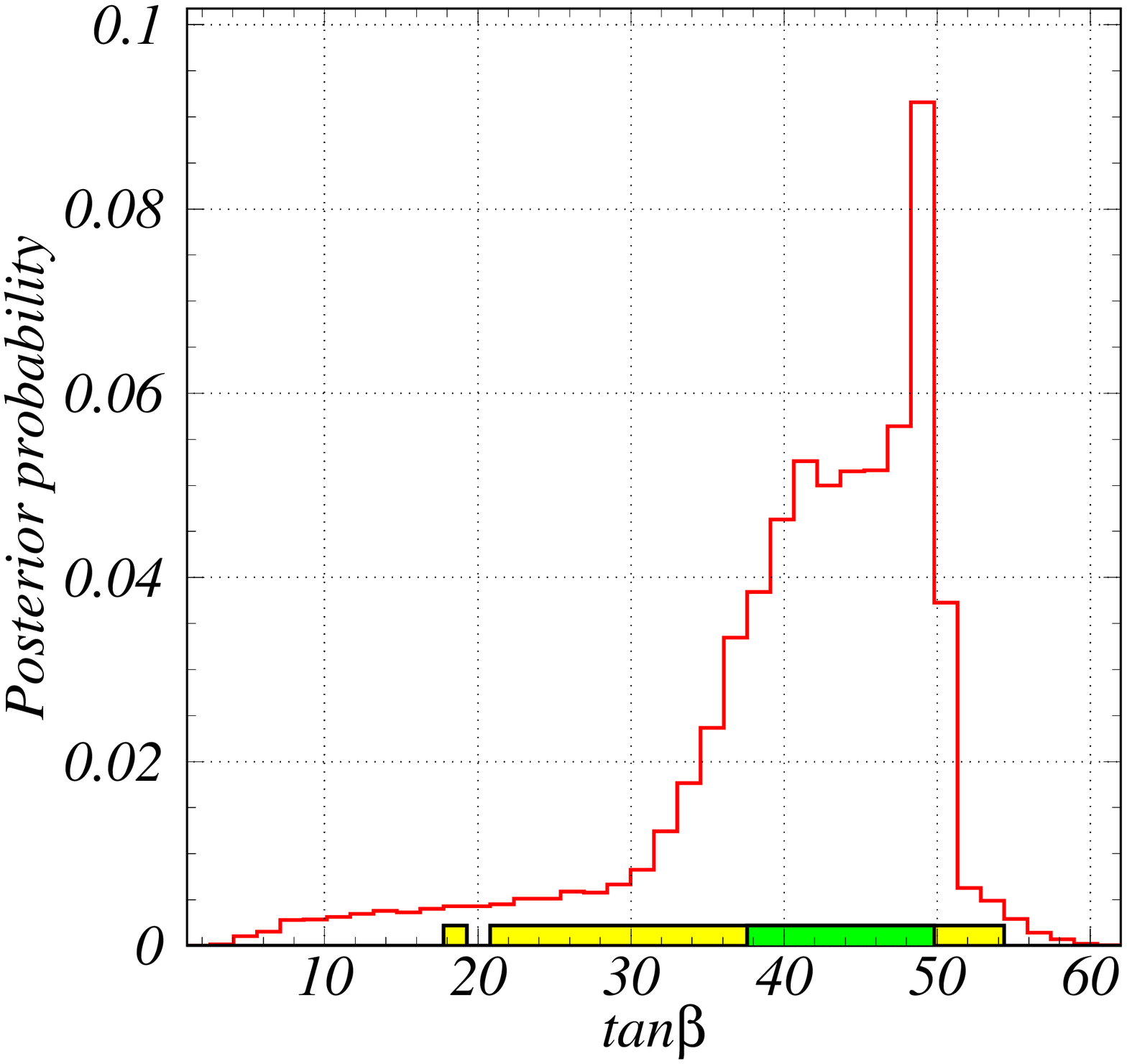,width=0.49\textwidth}
\caption{Posterior probability densities marginalized to the NmSuGra input parameters.}
\label{fig:MPP.inputs}}

The top left frame of Figure \ref{fig:MPP.inputs} shows the posterior probability marginalized to $M_0$.  As expected, based on the profile likelihood for $M_0$, the posterior peaks at low $M_0$ values.  But notably, it peaks at a higher value, around $M_0 \simeq 0.5$ TeV, than the likelihood function.  This difference is a good example of the volume effect: the sfermion-neutralino co-annihilation region requires a high fine-tuning between the sfermion (typically stau) and neutralino masses, and consequently it is extremely narrow in $M_0$.  Since it covers such a small region its posterior probability is suppressed even though its likelihood is high.

The exact opposite happens in the focus point region at high $M_0$.  The prominent 
rise of the posterior probability toward high $M_0$ values clearly comes from the overwhelming size of the focus point region.  Most of this enhancement happens at high $\tan\beta (\sim 50)$ where the traditional focus point region merges with multiple Higgs resonance corridors creating very wide regions consistent with WMAP.

The posterior probability distribution of $M_{1/2}$, in the top right frame of Figure \ref{fig:MPP.inputs}, develops two maxima.  A narrow peak is close to 150 GeV, and a wide one around 700 GeV.  In the dark matter context, the former corresponds to neutralinos resonantly self-annihilating via the lightest scalar Higgs boson in the s-channel.  This 'sweet spot' emerges as a combined high-likelihood and volume effect, as we will see a bit later.  Most of the wide peak comes from neutralinos resonance annihilating via the heavier scalar and pseudo-scalar Higgses.  These regions tend to create wide Higgs channels consistent with WMAP.

The same volume effect boosts the moderately positive $A_0$ region of the likelihood function in the posterior probability shown in the lower left frame of Figure \ref{fig:MPP.inputs}.  An even more prominent volume enhancement reshapes the posterior probability around $\tan\beta\sim 40-50$, as shown in the lower right frame.  The wide bump originates from Higgs resonance corridors, while the narrow peak from the focus point region which also raises the posterior at high $M_0$.

\FIGURE[th]{
\epsfig{file=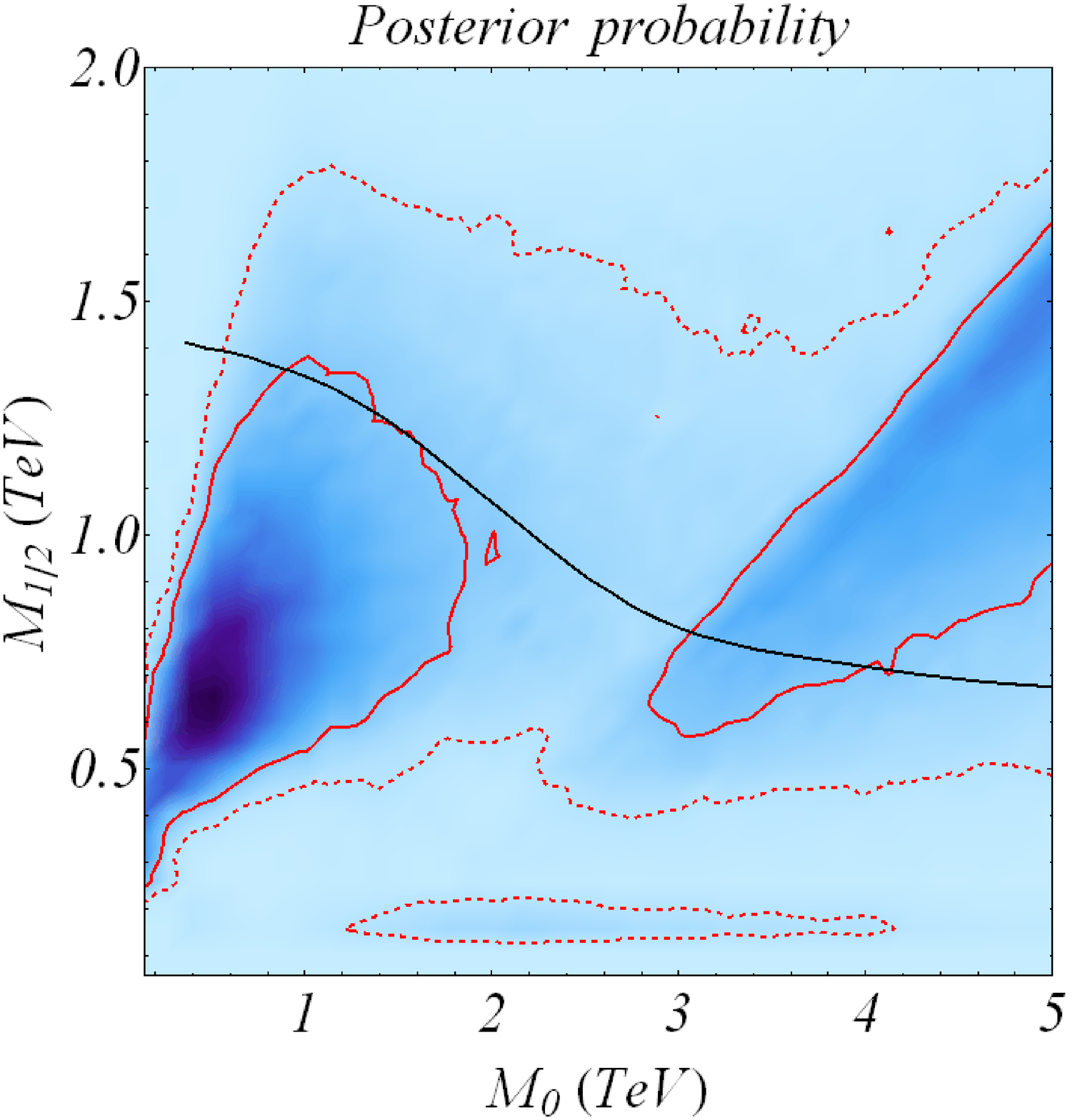,width=0.49\textwidth}
\epsfig{file=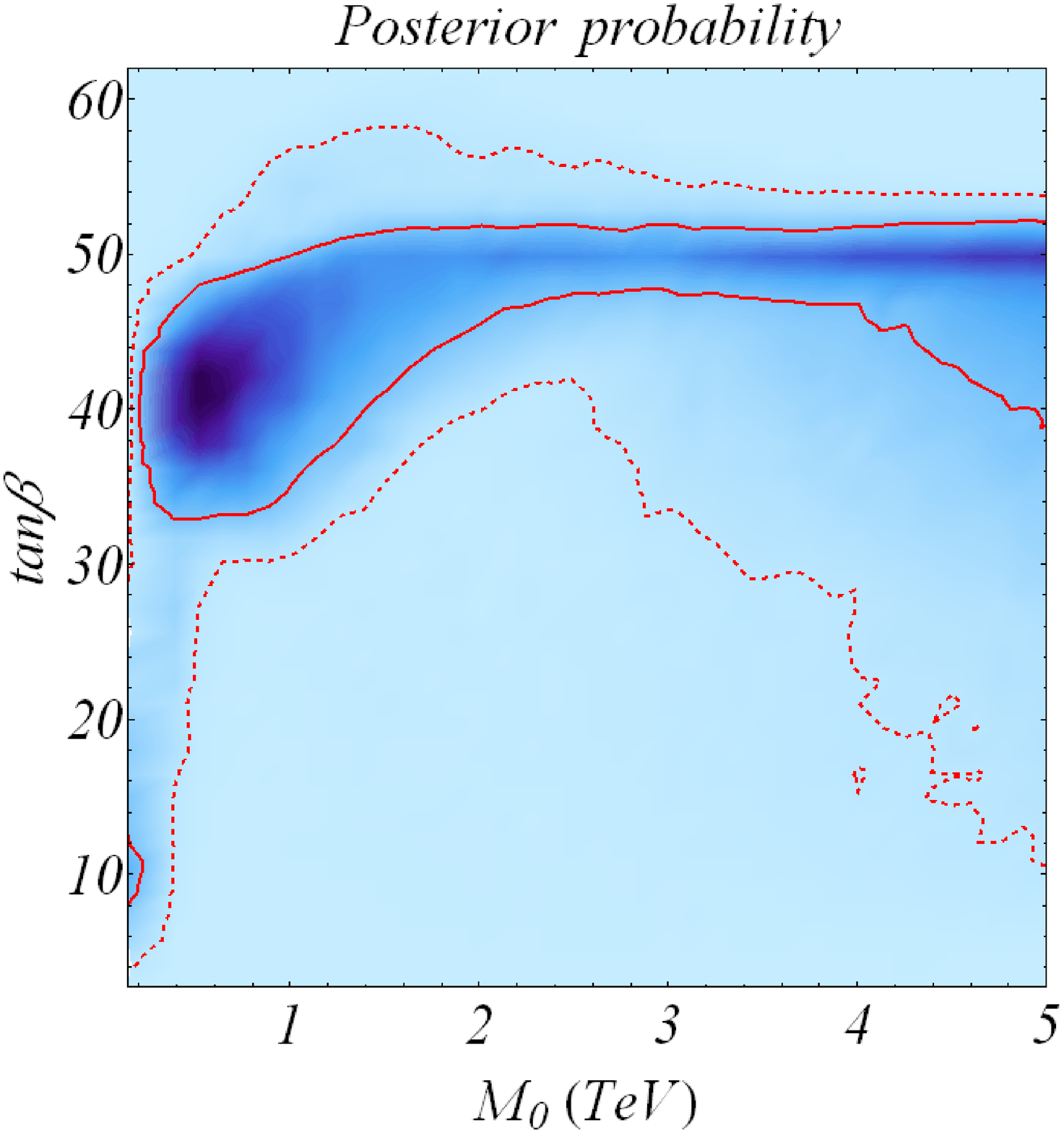,width=0.49\textwidth}
\epsfig{file=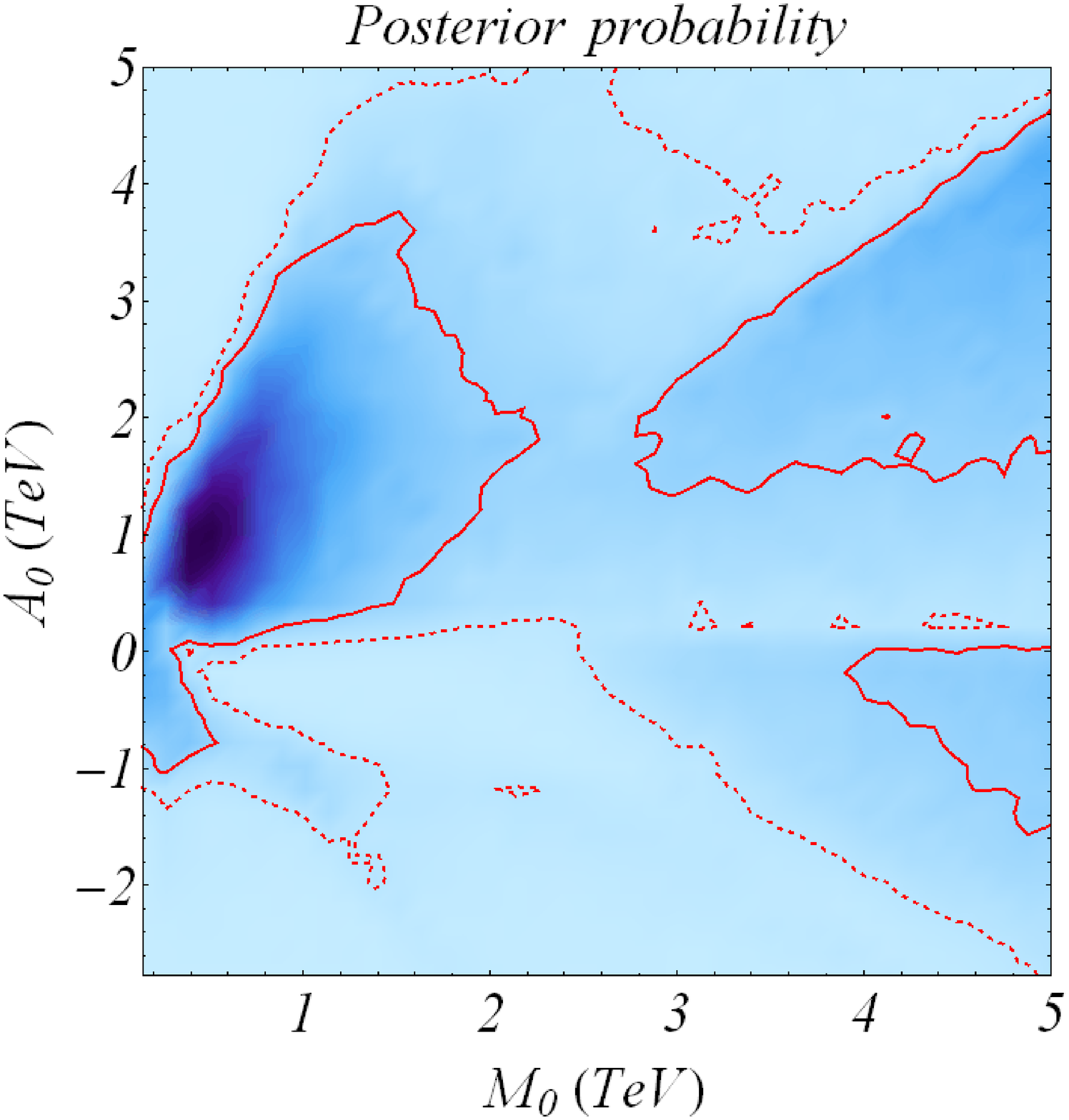,width=0.49\textwidth}
\epsfig{file=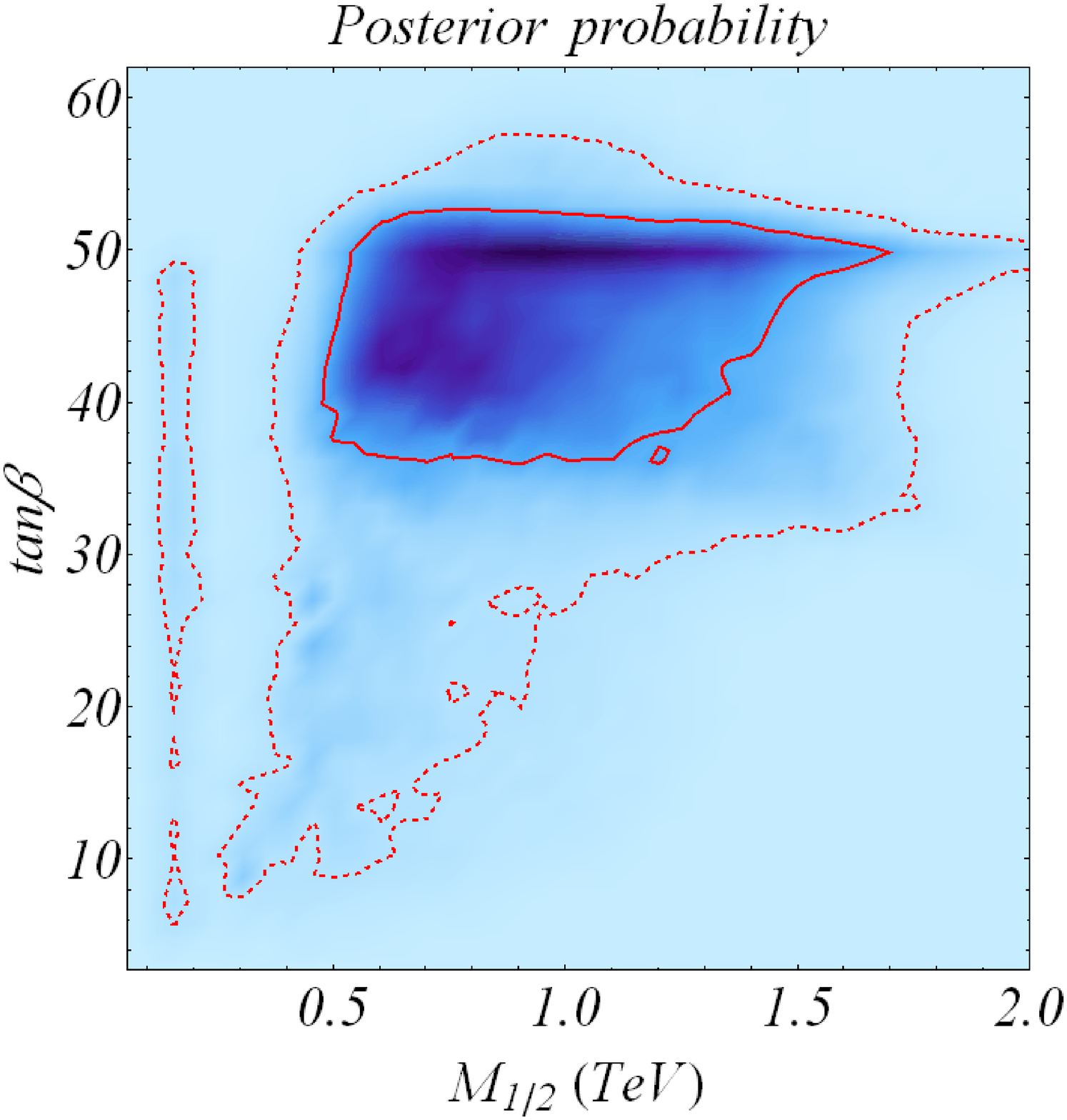,width=0.49\textwidth}
\caption{Posterior probability densities marginalized to pairs of NmSuGra input parameters.  The higher probability regions are darker.  Solid (dotted) red lines indicate 68 (95) percent confidence level contours.  On the top left frame the diagonal black curve shows the estimated reach of the LHC for 100 fb$^{-1}$ luminosity \cite{Baer:2003wx}.}
\label{fig:MPP2D.inputs}}

The effects described are evident from the posterior probability distributions marginalized to different pairs of NmSuGra input parameters, as Figure \ref{fig:MPP2D.inputs} shows.  In the top left frame we show the posterior probability marginalized to the plane of the common scalar and gaugino masses, $M_0$ vs. $M_{1/2}$.  The slepton co-annihilation region combined with Higgs resonance corridors, at low $M_0$ and low to moderate $M_{1/2}$ supports most of the probability.  This region is clearly separated from the focus point at high $M_0$ and moderate to high $M_{1/2}$, large part of which falls in the 68 \% confidence level.  

At the lowest $M_{1/2}$ values lies a Higgs annihilation strip, where the lightest neutralinos resonance annihilate via the lightest scalar Higgs boson.  Part of this region is allowed in NmSuGra due to the somewhat relaxed mass limit by LEP on the lightest Higgs.  The narrowness of this strip correlates with the smallness of the lightest Higgs width.  In this region the likelihood can be high, and when integrating over $M_0$ volume is accumulated.  This explains the narrow, isolated peak at the lowest values in the posterior probability marginalized to $M_{1/2}$.

The top right frame of Figure \ref{fig:MPP2D.inputs} shows the distribution of the posterior probability in the $M_0$ vs. $\tan\beta$ frame.  This makes it clear that most of the probable points are carried by Higgs resonant corridors toward higher $\tan\beta$, and the sfermion co-annihilation, due to its narrowness in $M_0$, falls only in the 95 \% confidence, but is outside the 68 \% region.  The exception is a minute corner of the parameter space at very low $M_0$, $M_{1/2}$, and $\tan\beta\sim 10$ where all theoretical results conspire to match experiment, raising the sfermion co-annihilation region into the 68 \% confidence region.  At the opposite, high $M_0$ and $\tan\beta$ corner multiple Higgs resonances combined with neutralino-chargino co-annihilation in the focus point lead to substantial contribution to the total probability.

The lower left frame of Figure \ref{fig:MPP2D.inputs} shows that positive values of $A_0$ are preferred over negative ones, because Higgs resonance annihilation occurs overwhelmingly at low to moderately positive values of $A_0$.  The focus point, containing less probability, extends further along both positive and negative $A_0$.  The lower right frame confirms that resonance annihilation via the lightest Higgs boson happens at all $\tan\beta$ values.  As previously mentioned, the resonance annihilation via heavier Higgs bosons and the focus point is only dominant at high $\tan\beta$.

\section{Experimental detection of NmSuGra}
\label{sec:detection}

Figure \ref{fig:PLD.inputs}, in itself, is encouraging for the prospects of discovery of NmSuGra at the LHC.  We can quantify this statement more precisely by constructing the profile likelihoods for the lightest superpartners and the gluino.  Figure \ref{fig:PLD.masses} shows that in the most likely NmSuGra model regions the mass of the lightest neutralino falls in the $50 \leq m_{\tilde{Z}_1} \lesssim 200~(375)$ GeV region at 68 (95) \% confidence level.  In these regions the lighter stau is relatively light, while the gluino and the lighter stop are moderately heavy.  

\FIGURE[th]{
\epsfig{file=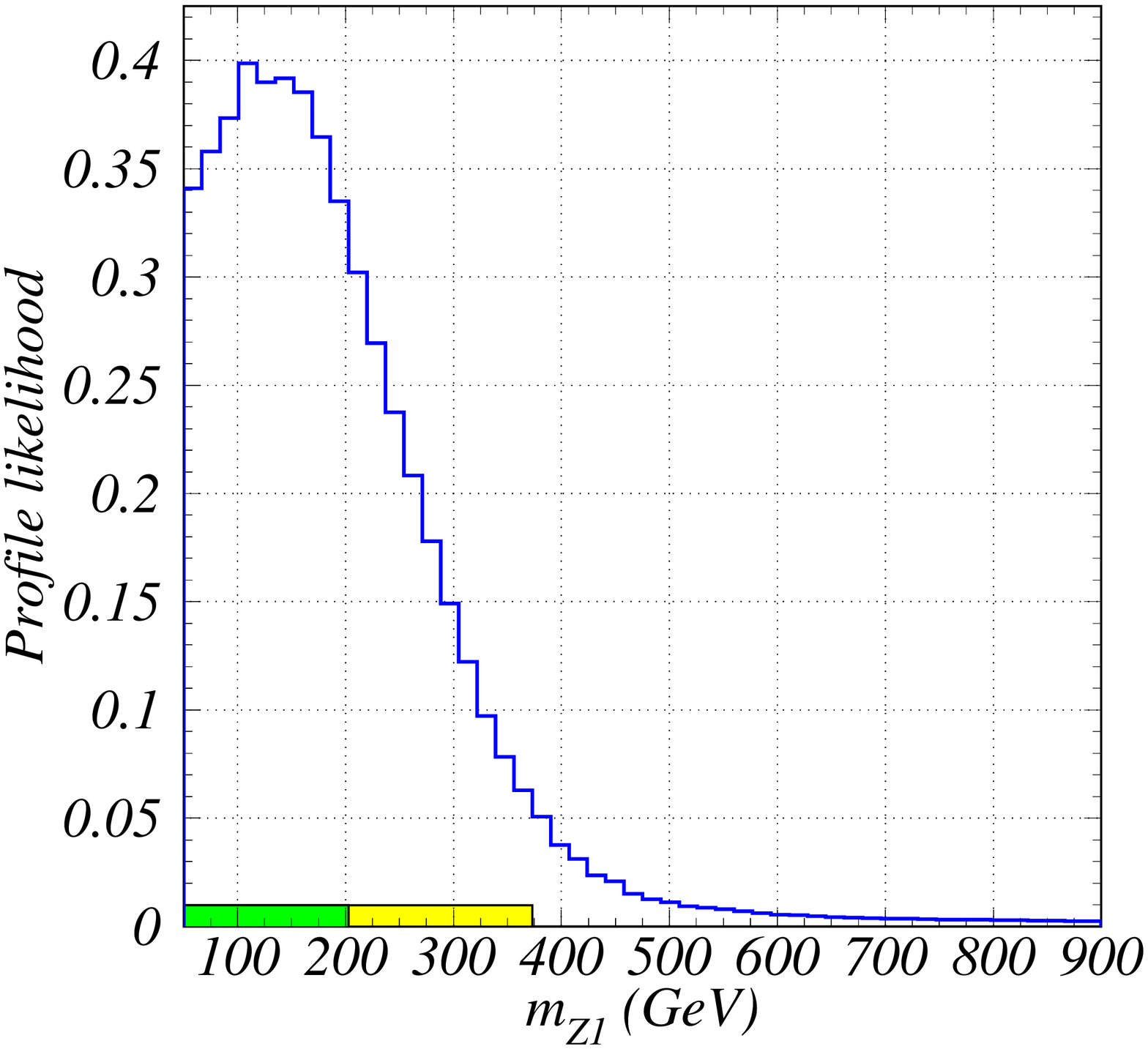,width=0.49\textwidth}
\epsfig{file=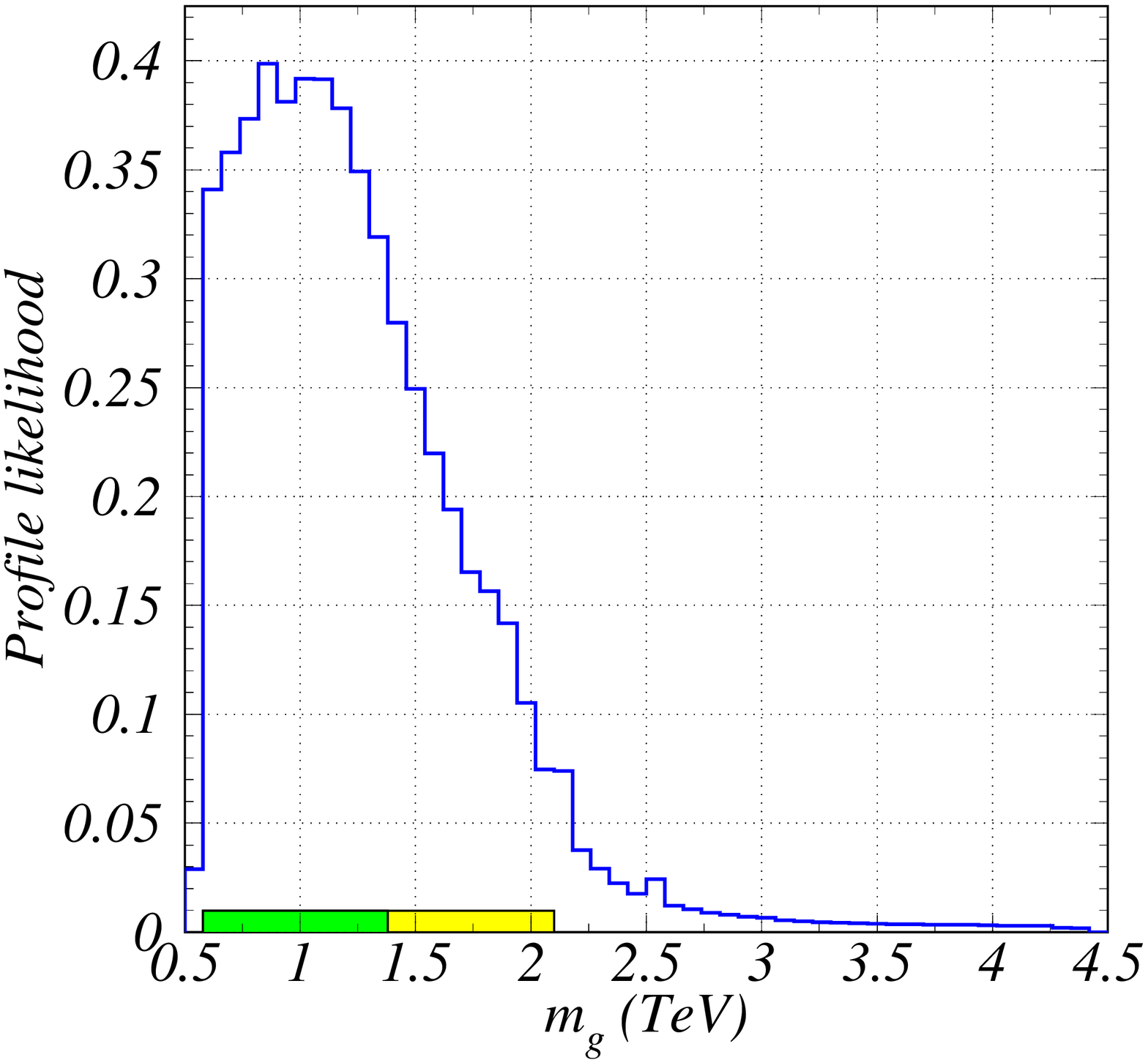,width=0.49\textwidth}
\epsfig{file=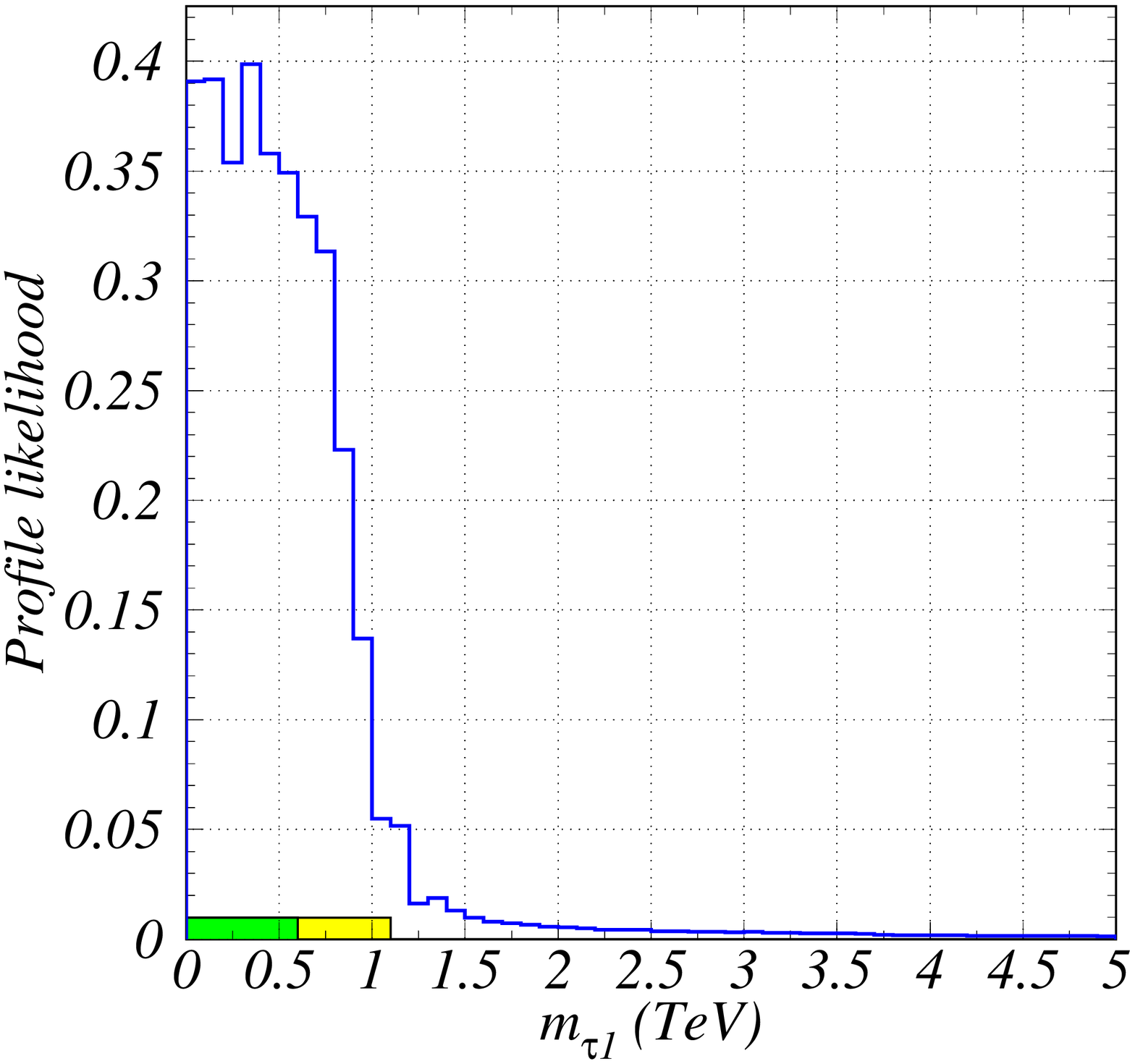,width=0.49\textwidth}
\epsfig{file=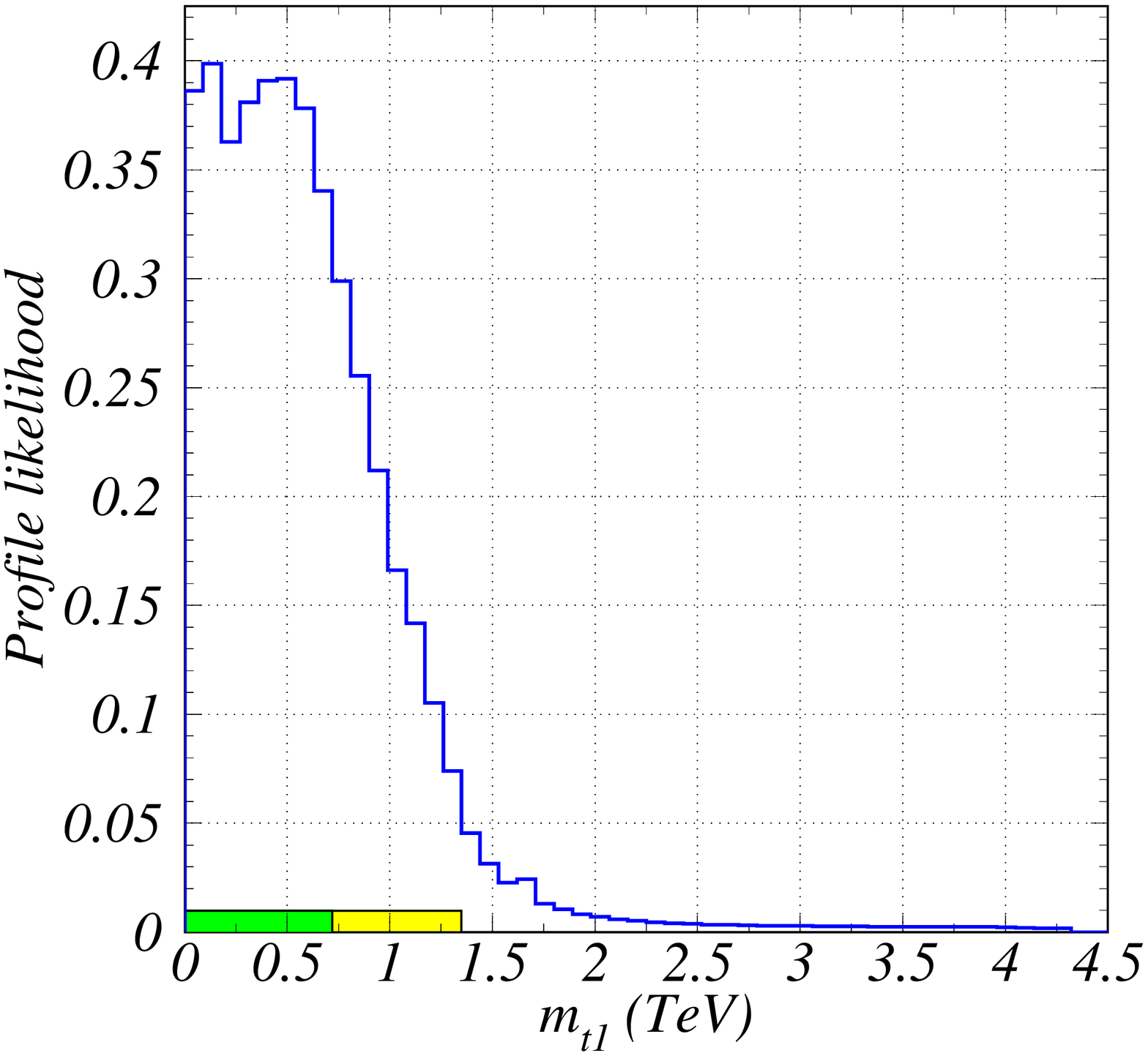,width=0.49\textwidth}
\caption{Profile likelihood distributions of various sparticle masses.  Green (yellow) coloring indicates 68 (95) percent confidence level regions.}
\label{fig:PLD.masses}}

The likelihood function of the lightest stop features a narrow peak at the lowest masses.  This tells us that high likelihoods are possible to achieve in NmSuGra parameter regions where the lightest stop co-annihilates with a neutralino, with both sparticle masses around 100 GeV.  This region, featuring low $M_0$, $M_{1/2}$, $\tan\beta$ and moderate $A_0$, is very interesting phenomenologically, since at these parameter values electroweak baryogenesis can solve the baryon asymmetry problem in the MSSM and in its singlet extensions \cite{Balazs:2004bu, Balazs:2004ae, Menon:2004wv, Balazs:2007pf}.  This region is excluded in mSuGra by the LEP Higgs mass limit, but in NmSuGra this limit is relaxed.  Thus, in the NmSuGra model a strongly first order electroweak phase transition might be made possible by a light stop, a moderate size singlet-Higgs coupling $\lambda$, or the combination of these.

After integrating the likelihood function over the considered NmSuGra parameter space, we can assess the detection prospects at the LHC.  The posterior probability in the top left frame of Figure \ref{fig:MPP.masses} indicates that the lightest neutralino mass is expected to lie in the $175 (125) \leq m_{\tilde{Z}_1} \lesssim 475 (675)$ GeV region at 68 (95) \% confidence level.  The 95 \% confidence level region also includes a small window around 70 GeV.  The shift toward heavier masses, relative to the likelihood functions, occurred due to heavier average sparticle masses in the focus point.  

\FIGURE[th]{
\epsfig{file=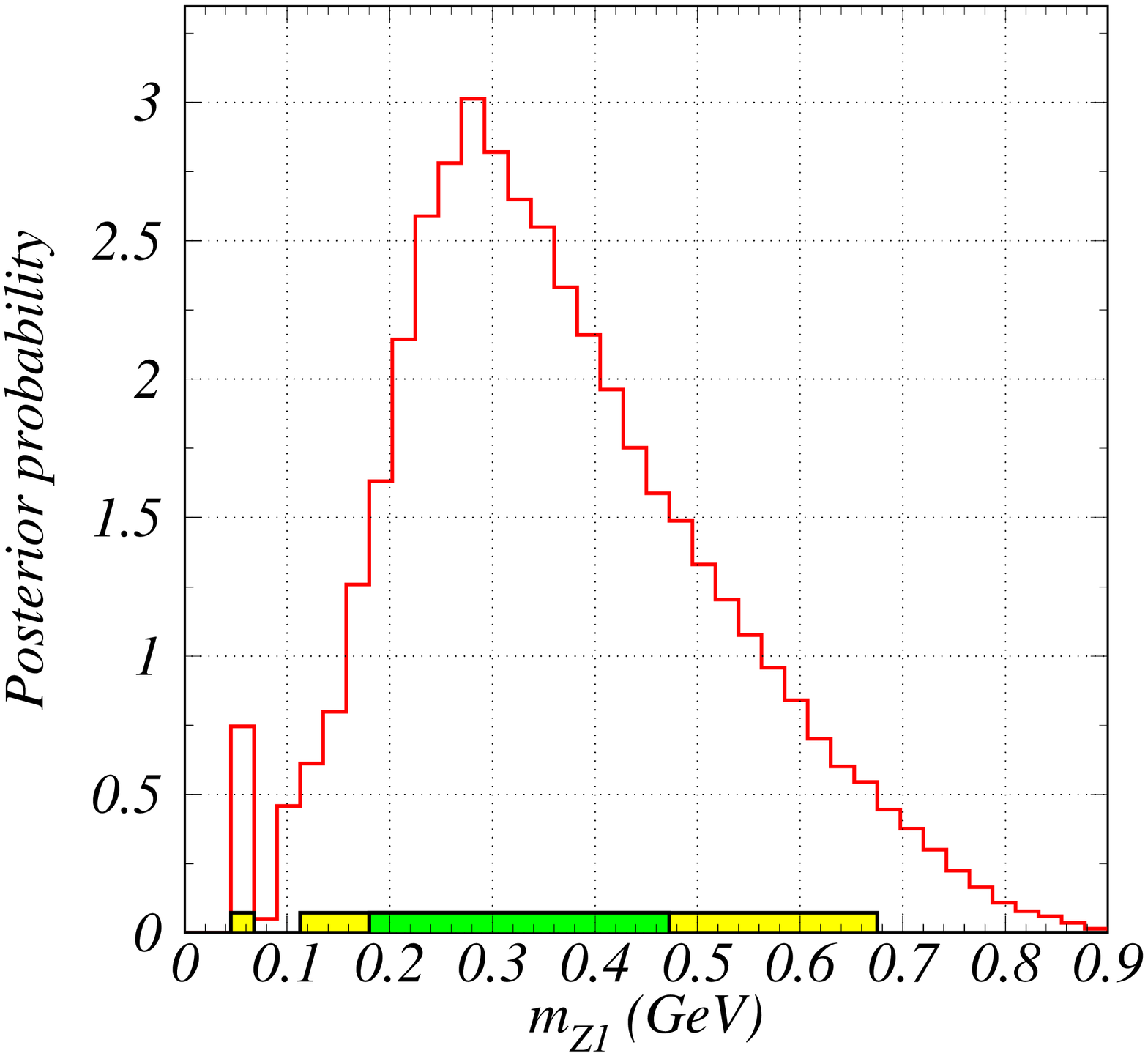,width=0.49\textwidth}
\epsfig{file=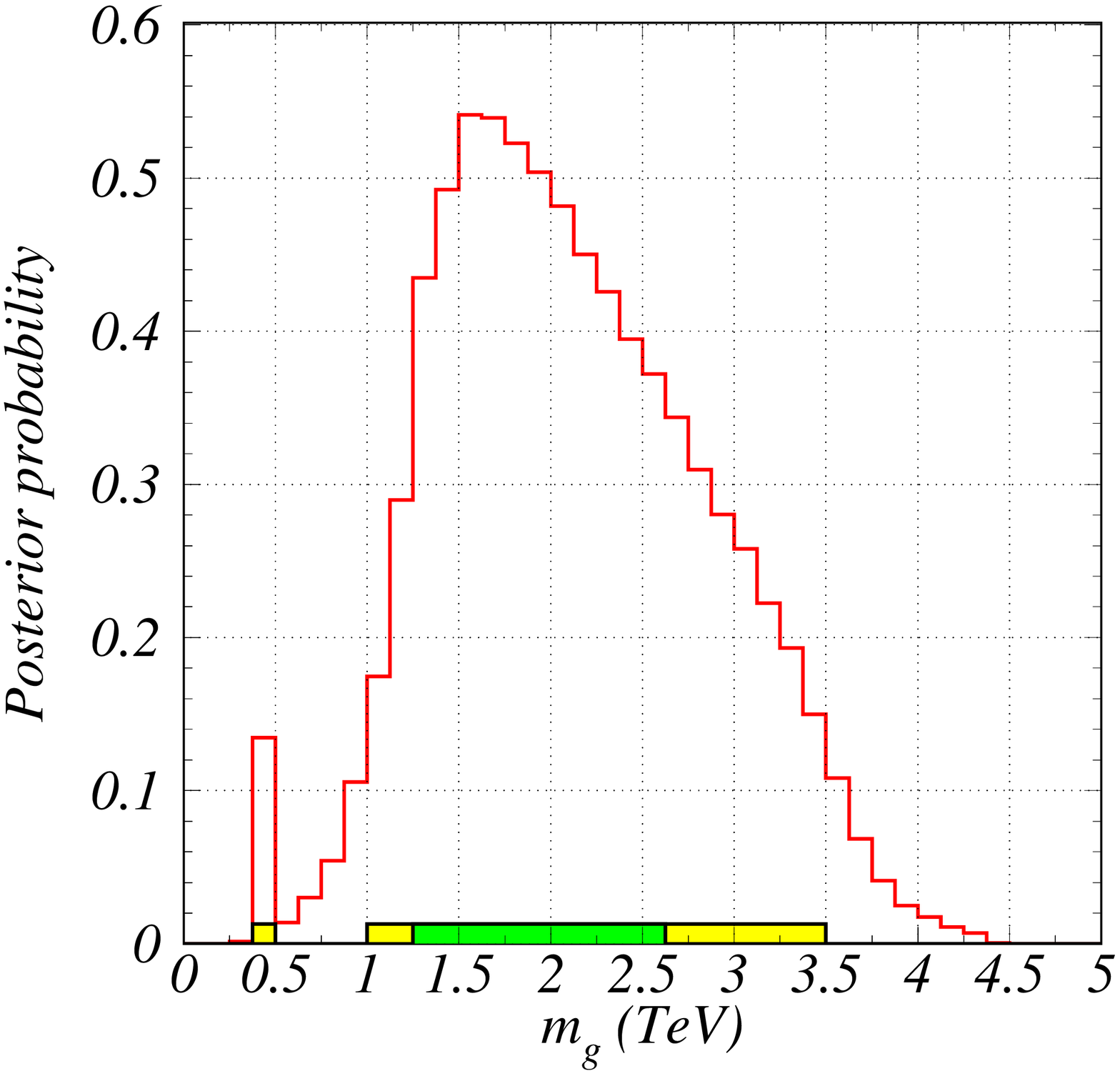,width=0.49\textwidth}
\epsfig{file=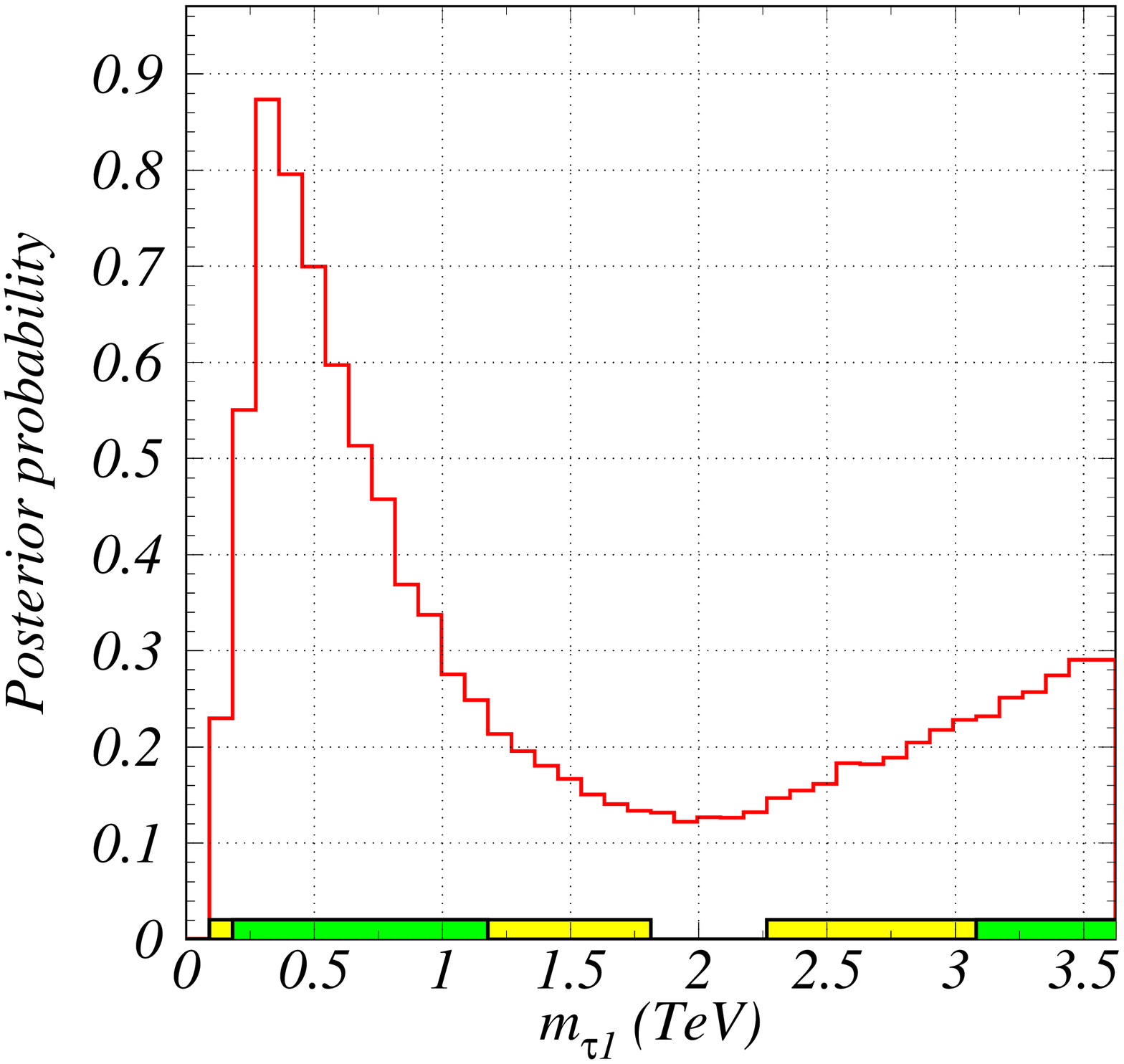,width=0.49\textwidth}
\epsfig{file=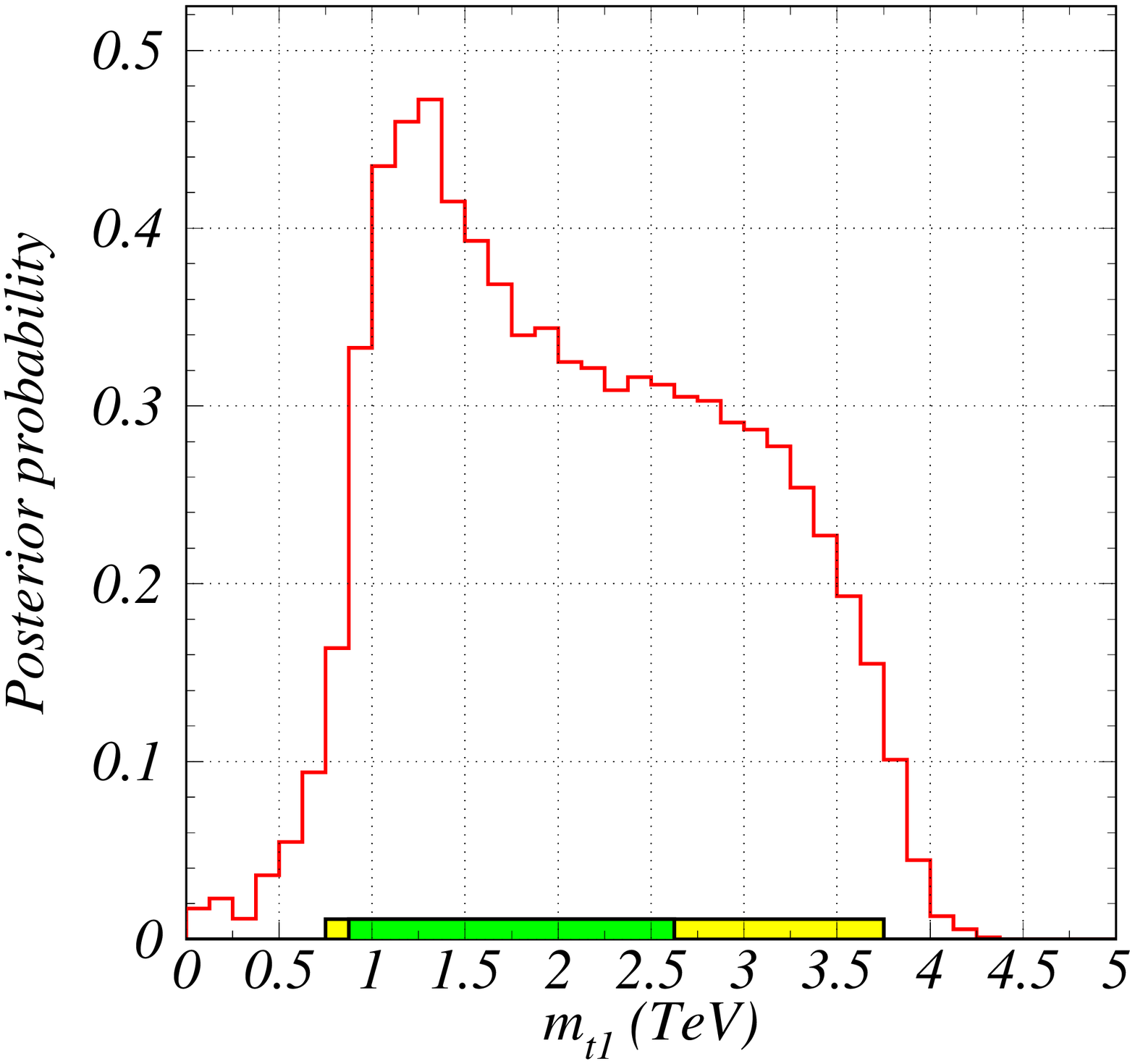,width=0.49\textwidth}
\caption{Posterior probability densities marginalized to various sparticle masses.}
\label{fig:MPP.masses}}

Unfortunately, this effect moves part of the NmSuGra parameter space out of the reach of the LHC, as shown by the posterior probability distribution of the gluino mass.  In the mSuGra model the LHC is able to reach about 3 TeV gluinos with 100 fb$^{-1}$ luminosity, provided the model has low $M_0$ \cite{Baer:2003wx}.  In the focus point this reach is reduced to about 1.75 TeV.  This means, just as in mSuGra, that part of the viable parameter space will remain out of reach of the LHC.

This conclusion is quantified even better by the estimated LHC reach displayed in the top left frame of Figure \ref{fig:MPP.inputs}.  From there it is evident that with 100 fb$^{-1}$ the LHC will be able to cover sfermion co-annihilations and Higgs resonances that fall into the 68 \% confidence region, together with a small part of the focus point.  But the LHC will stop short of fully exploring the Higgs resonances and the focus point at high $M_{1/2}$.

In the lower left frame of Figure \ref{fig:MPP.masses} the posterior probability distribution of the lightest stau mass mirrors that of $M_0$.  The next frame shows that the lighter stop is also expected to be heavier than the likelihood function alone suggests.  Even the sharp peak at low values in the stop likelihood function is overwhelmed due to the minute volume of the parameter space it occupies.

\FIGURE[th]{
\epsfig{file=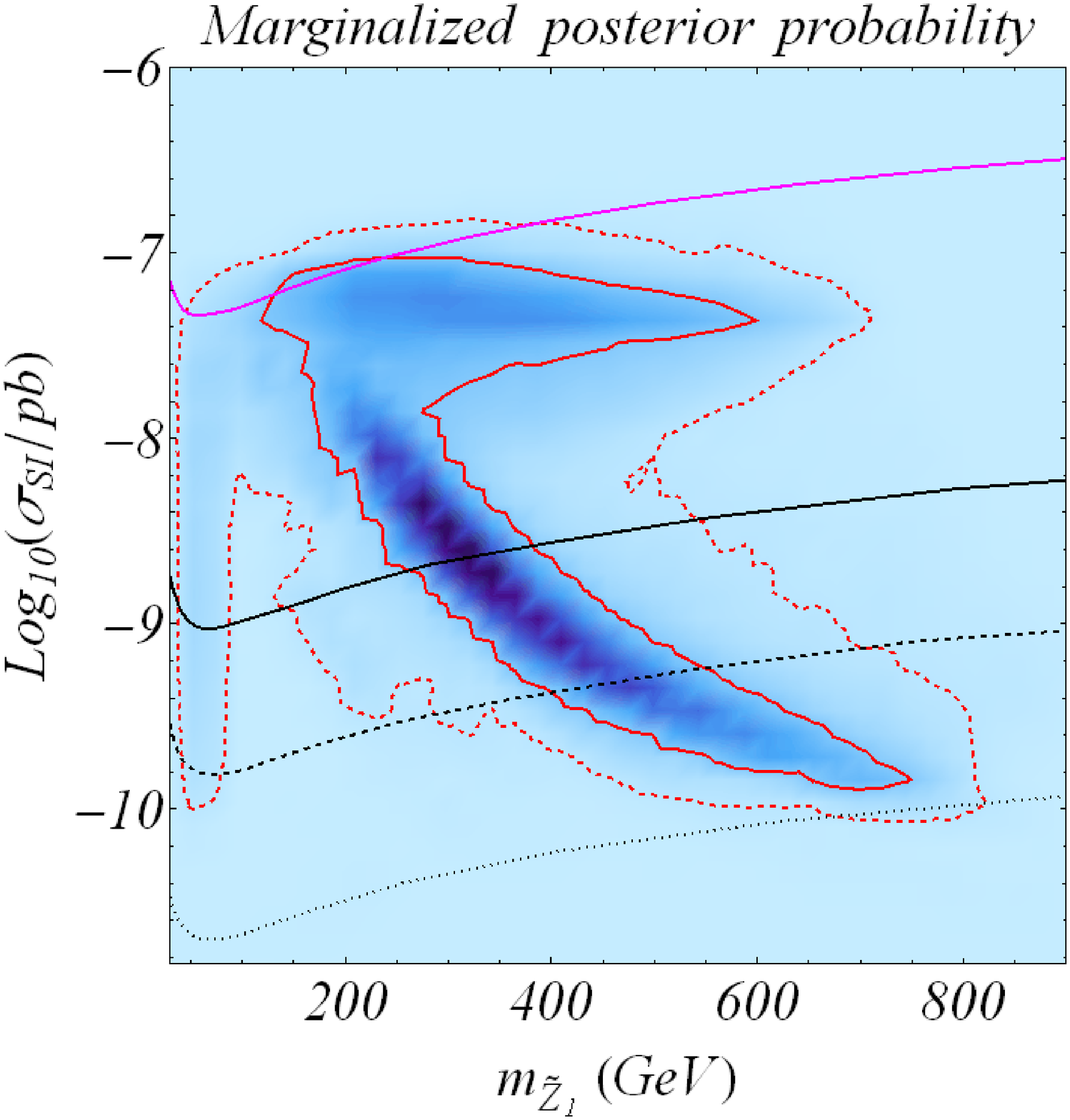, width=0.49\textwidth}
\caption{Posterior probability density marginalized to the spin-independent neutralino-nucleon elastic recoil cross section and the lightest neutralino mass.  Confidence level contours are shown for 68 (solid red) and 95 (dashed red) \%.  The present (solid magenta) and projected reach of the upgraded CDMS experiment is shown for a 25 (solid black), 100 (dashed black), and a 1000 (dotted black) kg detector.}
\label{fig:MPP.ssi}}

While the LHC will not be able to cover the full viable NmSuGra parameter space, fortunately measurements in the near future will explore a large part of the remaining region.  Amongst the most promising experiments complementing the capabilities of the LHC are the measurements of the spin-independent neutralino-nucleon elastic recoil cross section, $\sigma_{SI}$.  From several of these experiments, we single out CDMS as the most illustrative example.  Figure \ref{fig:MPP.ssi} shows the posterior probability density marginalized to the plane of $\sigma_{SI}$ and the lightest neutralino mass.

The magenta line toward the top of the plot is the present upper limit set by CDMS in 2008 \cite{Ahmed:2008eu}.  Although this limit is included in our likelihood function, due to our generous estimate of the theoretical error (20\%), the combined theoretical-experimental likelihood function still allows a small region above the experimental exclusion.  The black line between $10^{-9} < \sigma_{SI} < 10^{-8}$ shows the estimated reach of a 25 kg super-CDMS.  The lower dashed and dotted black lines show the estimated CDMS reach for a 100 and 1000 kg detector \cite{Akerib:2006rr}.

This plot clearly shows that direct detection experiments, if performs as expected, can play a pivotal role in discovering or ruling out simple constrained supersymmetric scenarios.  It is interesting to observe that a one ton version of CDMS alone would be able to cover the full relevant NmSuGra parameter space.  But the most important message is that even a 25 kg CDMS will reach a substantial part of the focus point region, complementing the LHC.  

In the posession of the above results, we can quantify the chances for the discovery of NmSuGra at the LHC by calculating the ratio of posterior probabilities with and without the LHC reach:
\begin{eqnarray}
 \frac{\int_{\rm within~LHC~reach} \mathcal{P}(p_i|D;H) dp_i}
      {\int_{\rm outside~LHC~reach} \mathcal{P}(p_i|D;H) dp_i} = 0.57 .
\end{eqnarray}
{\it According to this the odds of finding NmSuGra at the LHC are 4:3} (assuming, of course, that the model is chosen by Nature).
We can also easily calculate the probability of the discovery of NmSuGra at the LHC combined with a ton equivalent of CDMS (CDMS1T):
\begin{eqnarray}
 \frac{\int_{\rm within~LHC+CDMS1T~reach} \mathcal{P}(p_i|D;H) dp_i}
      {\int_{\rm outside~LHC+CDMS1T~reach} \mathcal{P}(p_i|D;H) dp_i} = 0.99 .
\end{eqnarray}
This means that according to the present data the NmSuGra model lies within the combined reach of the LHC and CDMS1T at 99 percent confidence level.  The two experiments combined are essentially guaranteed to discover this model!  This result strongly underlines the complementarity of collider and direct dark matter searches.

\section{Conclusions}

The next-to-minimal supergravity motivated model is one of the more compelling models for physics beyond the standard model due to its naturalness and simplicity.  In this work we applied a thorough statistical analysis to NmSuGra based on numerical comparisons with present experimental data.  Using Bayesian inference we found that significant regions of the NmSuGra parameter space remain viable under current constraints, and that the LHC has a 57 \% chance to discover 
these regions.  Furthermore, we found that the predicted LHC reach combined with the projected sensitivity of a ton equivalent of CDMS covers the viable NmSuGra parameter region at 99 \% confidence level.  This result underlines the complementarity of the LHC and direct dark matter searches in discovering new physics at the TeV scale.  

Since much of the NmSuGra phenomenology appears to be very similar to that of the MSSM after imposing minimal supergravity (mSuGra or CMSSM), we expect these conclusions to be broadly valid within the constrained MSSM as well.  While this is good news from the theoretical viewpoint, it poses a challenge to the LHC experimentalists to disentangle these models.

\section{Acknowledgments}

The authors are indebted to D. Kahawala, F. Wang, L. Roszkowski, and M. White for invaluable discussions on various aspects of the NMSSM and the likelihood analysis.  This research was funded in part by the Australian Research Council under Project ID DP0877916.


\bibliographystyle{JHEP}
\bibliography{paper}


\end{document}